\documentclass[aps,twocolumn,prx,floatfix,superscriptaddress,showpacs,longbilbiography]{revtex4-1}
\usepackage[utf8]{inputenc} 

\usepackage{graphicx}
\usepackage{dcolumn}
\usepackage{bm}
\usepackage{enumerate}

\usepackage{braket}
\usepackage{amsfonts}
\usepackage{amsmath}
\usepackage{amssymb}

\usepackage{color,soul}
\usepackage{wasysym}
\usepackage{mathrsfs}
\usepackage{times}

\usepackage[dvipsnames,svgnames,table]{xcolor}
\usepackage{hyperref}
\usepackage{fancyhdr}
\usepackage{float}
\usepackage[english]{babel}
\usepackage[caption=false]{subfig}
\usepackage{braket}

\usepackage{graphicx}
\hypersetup{
  breaklinks = {true},
  citecolor = {blue},
  colorlinks = {true},
  linkcolor = {red},
}
\usepackage{physics}
\usepackage{mathtools}
\usepackage[dvipsnames,svgnames]{xcolor}
\usepackage{braket} 
\usepackage{upgreek}

\usepackage[normalem]{ulem} 

\hypersetup{
pdfstartview={FitH},
colorlinks=true,    
linkcolor=NavyBlue, 
citecolor=Maroon,   
filecolor=NavyBlue, 
urlcolor=NavyBlue   
}

\newcommand{\bea}{\begin{eqnarray}}
\newcommand{\eea}{\end{eqnarray}}

\renewcommand{\selectlanguage}[1]{}

\begin{document}
\title{A non-Hermitian loop for a quantum measurement}

\author{Luis E. F. {Foa Torres}$^*$}
\affiliation{Departamento de F\'{\i}sica, Facultad de Ciencias F\'{\i}sicas y Matem\'aticas, Universidad de Chile, Santiago, 8370415, Chile}
\email{e-mail: luis.foatorres@uchile.cl}
\author{Stephan Roche}
\affiliation{Catalan Institute of Nanoscience and Nanotechnology (ICN2), CSIC and BIST, Campus UAB, Bellaterra, 08193 Barcelona, Spain}
\affiliation{ICREA--Instituci\'o Catalana de Recerca i Estudis Avan\c{c}ats, 08010 Barcelona, Spain}

\begin{abstract}

Here we present a non-Hermitian framework for modeling state-vector collapse under unified dynamics described by Schrödinger's equation.
Under the premise of non-Hermitian Hamiltonian dynamics, we argue that collapse has to occur when the Hamiltonian completes a closed loop in the parameter space encoding the interaction with the meter. For two-level systems, we put forward the phenomenon of chiral state conversion as a mechanism for effectively eliminating superpositions. This perspective opens a way to simulate quantum measurements in classical systems that up to now were restricted to the Schrödinger part of the quantum dynamics.
\end{abstract}

\date{\today}
\maketitle

\section{Introduction} 

One century after its inception, quantum mechanics stands as one of humanity's most rigorously tested theories, yet it still keeps a foundational challenge known as the measurement problem~\cite{bell_against_1990,dirac_evolution_1963,laloe_we_2001}. The crux of the measurement problem lies in explaining or making sense of the wave function collapse. The postulates of quantum mechanics state that during a measurement, the Schrödinger equation, which ordinarily governs quantum dynamics, is momentarily suspended to allow the wave function to collapse (see, for example,~\cite{cohen-tannoudji_quantum_2020,shankar_principles_1994}). This split description of the quantum dynamics is introduced to bridge the complex world of quantum superpositions with the familiar classical reality, where those superpositions have vanished. The incompatibility between the two descriptions for the dynamics entails a \textit{``cut"} between the quantum and classical realms (as Werner Heisenberg called it in a posthumously published letter~\cite{bacciagaluppi_translation_2011} in response to the EPR paper~\cite{einstein_can_1935}). Indeed, the collapse is essentially a non-unitary, irreversible operation with loss of information and eventual energy and momentum exchange~\cite{pearle_wavefunction_2000}, or as John Wheeler put it \textit{``an irreversible act of amplification such as the blackening of a grain of silver bromide emulsion"}~\cite{wheeler_law_2016}~\footnote{Wheeler mentions that the term \textit{irreversible amplification} was used by Bohr, see chapter 1 of~\cite{bohr_atomic_2010}, which was published in 1958.}.

Despite advances in understanding the quantum-classical transition~\cite{zurek_decoherence_2003,schlosshauer_m_decoherence_2007}, the measurement problem remains a mystery for many of us~\cite{leggett_problems_2006,hance_what_2022}~\footnote{There are also notable opinions against thinking of the measurement problem as such, see for example~\cite{mermin_there_2022}. We also refer to~\cite{schlosshauer_measurement_2011}
for a broader overview of the different positions}. Two proposals to address the conventional view of measurements are the Ghirardi-Rimini-Weber (GRW)~\cite{ghirardi_unified_1986,ghirardi_markov_1990} and the Continuous Spontaneous Localization (CSL) models~\cite{pearle_combining_1989}. Both introduce stochastic elements to the Schrödinger's equation, suggesting that wave function collapse can occur spontaneously~\cite{bassi_dynamical_2003} (there are also other proposals using stochastic elements such as~\cite{fernandez-alcazar_decoherent_2015}). Although GRW models do not rely on the concept of an apparatus to induce collapse, they still lack a microscopic explanation for how collapse occurs. Other proposals, inspired by GRW, are due to Diósi~\cite{diosi_models_1989,donadi_underground_2021} and Penrose~\cite{penrose_gravitys_1996} linking the wave function collapse to gravity. More recently, different authors (see ~\cite{martinez_romeral_wavefunction_2024,singh_embedding_2023,singh_emulating_2023} and also~\cite{van_wezel_broken_2010,mertens_objective_2023,mukherjee_colored-noise-driven_2024}) proposed using a non-Hermitian Hamiltonian to model a quantum measurement (earlier proposals also include dissipative modifications of the Schrödinger equation~\cite{gisin_collapse_1981}). These models are steps in making sense of quantum collapse~\cite{bell_against_1990}, yet they also underscore its complexity~\cite{bassi_models_2013} and the need for further experiments.

In a parallel and unfolding story, over recent years, we have seen a rise in research fronts using photonics~\cite{rechtsman_photonic_2013,vicencio_observation_2015}, metamaterials~\cite{veenstra_non-reciprocal_2024}, acoustics~\cite{zhang_acoustic_2021}, mechanical systems~\cite{lorenz_classical_2023} and circuits~\cite{lee_topolectrical_2018} as classical analogs for quantum systems. Notably, in photonic lattices, the paraxial Helmholtz equation for the electromagnetic waves maps to the Schrödinger equation~\cite{szameit_discrete_2010} thus allowing to simulate quantum systems. Similarly, in electrical circuits, the correspondence between the circuit Laplacian and the tight-binding Hamiltonian have been extensively used to investigate topological states~\cite{lee_topolectrical_2018}. Moreover, it has been demonstrated that these circuits can also mimic Schrödinger dynamics~\cite{ezawa_electric-circuit_2019}. Collectively, these examples illustrate the availability of a classical toolkit for simulation of quantum systems. 
However, a critical limitation remains: while a few noteworthy exceptions offer solutions for specific cases~\cite{longhi_nonexponential_2006,liu_engineering_2023}, classical simulators generally lack a broadly applicable counterpart to quantum measurements, thereby preventing the exploration of a wealth of phenomena such as measurement-enriched phases.

The aforementioned limitation of classical simulators in handling quantum measurement shows the need for a deeper understanding of the measurement process itself, which might then inspire new simulation approaches. Focusing on recent proposals modeling collapse using non-Hermitian Hamiltonians, challenges remain. Although models like our prior work~\cite{martinez_romeral_wavefunction_2024} show how non-Hermitian dynamics can amplify a chosen state, the need to construct the interaction specifically to favor a predetermined outcome perhaps offers less insight into the measurement process itself. Moreover, such approaches do not necessarily provide a dynamical explanation for \textit{why} the possible outcomes of an energy measurement correspond specifically to the eigenstates of the system's original Hamiltonian ($\mathcal{H}_0$). One might desire a mechanism where both the preference for eigenstates and the selection among them emerge dynamically from the system's response to control or interaction. This motivates our exploration of an alternative framework centered on dynamically induced collapse via parameter loops around exceptional points, aiming to demonstrate how such dynamics can naturally establish the initial eigenstates as the unique stable outcomes and provide a topological mechanism for selecting one.

The alternative framework we propose leverages key features of non-Hermitian Hamiltonians~\cite{moiseyev_non-hermitian_2011,bergholtz_exceptional_2021,lin_topological_2023}, systems whose unique properties offer a potential pathway to address the limitations discussed above. Non-Hermitian Hamiltonians exhibit phenomena absent in their Hermitian counterparts, most notably non-Hermitian degeneracies known as exceptional points (EPs)~\cite{kato_perturbation_1995,moiseyev_non-hermitian_2011}, where both eigenvalues and eigenvectors coalesce. These EPs are central to many counterintuitive phenomena~\cite{heiss_chirality_2001,dembowski_experimental_2001,heiss_physics_2012} and useful effects, including sensitivity enhancements relevant for novel sensors~\cite{wiersig_enhancing_2014,rechtsman_optical_2017,hodaei_enhanced_2017,chen_exceptional_2017,hokmabadi_non-hermitian_2019}. Crucially for our goal of dynamically explaining state selection, the behavior near EPs, particularly the state evolution during parameter loops encircling them~\cite{hassan_dynamically_2017}, forms the core of our proposed collapse mechanism. Furthermore, the physics of non-Hermitian systems and EPs can be readily explored in various experimental platforms, including metamaterials~\cite{veenstra_non-reciprocal_2024}, photonics~\cite{szameit_pt-symmetry_2011,pan_photonic_2018,miri_exceptional_2019,vicencio_nonsymmetric_2025}, acoustics~\cite{zhang_acoustic_2021}, and circuits~\cite{schindler_experimental_2011,hofmann_reciprocal_2020}. This experimental accessibility offers a unique opportunity not only to test the fundamental dynamics we propose but also potentially to simulate aspects of quantum collapse, as modeled here, within controllable classical systems.

To set the stage, we first revisit the postulates of quantum mechanics, laying the groundwork for our discussion. Subsequently, we discuss a set of physically motivated assumptions and their consequences. The main outcome of this part is that if a quantum measurement is to be modeled as a process that occurs in time and which is described by the Schrödinger equation with a non-Hermitian Hamiltonian encoding the effect of the meter (or whatever interaction leading to collapse), then the Hamiltonian should describe a closed loop in parameter space (returning to its initial Hermitian form). Interestingly, we are able to readily link this discovery with the phenomenon called state exchange or chiral state conversion~\cite{hassan_dynamically_2017}: For the $2\times 2$ case, when moving along a closed contour in the space of the Hamiltonian parameters, and under certain conditions for the speed with which the loop is completed, the initial state undergoes a conversion to a preferred state, which is determined by the sense of rotation in the space of parameters~\cite{hassan_dynamically_2017}. Thus, chiral state conversion breaks the superpositions and models quantum collapse under a single dynamic equation. 
However, there is a twist to consider: the sense of the loop is a property of the interaction itself rather than the initial Hermitian Hamiltonian. In other words, it is an additional variable, and compatibility with standard quantum mechanics requires that its value must be set to satisfy Born's rule.
We close by discussing the implications of our results on the path to simulating quantum measurement in classical systems.

\section{Split procedures for the quantum dynamics}
Quantum mechanics is traditionally based on four key postulates, as outlined in textbooks~\cite{cohen-tannoudji_quantum_2020,shankar_principles_1994}. Two of these postulates specifically address the theory's dynamics. One of them pertains to the standard, measurement-free behavior governed by the Schrödinger equation, highlighting how systems evolve when not being measured. The other postulate introduces the concept of measurement as a distinct phenomenon, where the act of measuring disrupts the usual dynamics, causing the system's state vector to \textit{collapse} into one of the observable's eigenstates. This collapse results in the system adopting a state that corresponds to one of the observable's eigenvalues, rendering the measurement as an exceptional process within quantum mechanics, described by a different antagonistic procedure.

The measurement postulate in quantum mechanics, in its conventional form, can be broken down into three key questions: 1) What are the possible outcomes of a measurement? 2) What happens to the system's state after a measurement (state-vector collapse)? and 3) What are the probabilities of obtaining each outcome (Born's rule)? Here, we primarily address the first and second question, offering a model for the dynamics of the collapse itself.

 \begin{figure*}
    \centering
\includegraphics[width=0.95\textwidth]{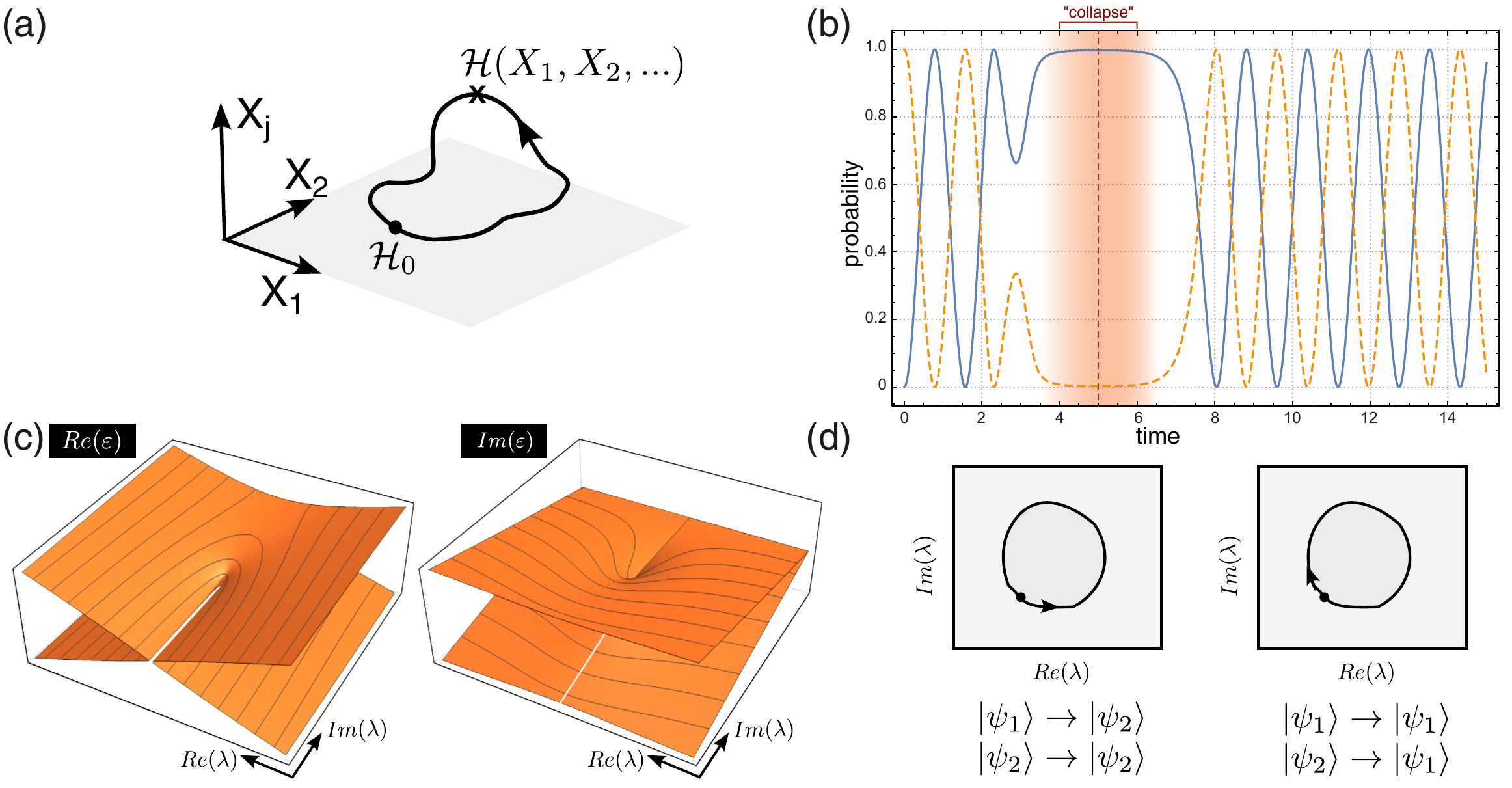}
    \caption{(a) Scheme showing a loop in a multidimensional parameter space defining the Hamiltonian. ${\cal H}_0$ is the initial Hamiltonian just before the interaction that leads to the collapse starts. (b) Time-evolution of the spin-up (blue, continous) and spin-down (orange, dashed) relative probabilities for a system evolving under the Hamiltonian of Eq.~ \ref{eq-H} with $\gamma_0=1$ and $\Gamma=50$. The state at $t=0$ corresponds to a particle with spin down. A non-Hermitian term adding gains on the spin-up and loss on the spin-down is turned on with a gaussian function centered at time $t=5$ (standard deviation is $1$). This pulsed perturbation leads to a collapse after which the time-evolution restarts. (c) Scheme showing the Riemann surfaces of the real and imaginary part of the eigenvalues near an exceptional point. (d) Because of the peculiar structure of the solutions in presence of an exceptional point, the phenomenon of chiral state conversion leads to the asymmetric conversion of the initial eigenstates $\ket{\psi_1}$ and $\ket{\psi_2}$ after a closed loop which defies the adiabatic intuition and destroys superposition states.}
    \label{fig1}
\end{figure*}

\textit{Dynamical Model of a Quantum Measurement.--} Our model for a dynamical description of quantum measurement using non-Hermitian Hamiltonians rests on three key assumptions:

\begin{enumerate}
    \item \textbf{Finite-Time Evolution:} The measurement occurs over a finite time interval. The system's state vector, denoted as \(|\Psi(t)\rangle\), evolves according to the standard linear time-dependent Schr\"odinger equation:
\[ i\hbar \frac{\partial |\Psi(t)\rangle}{\partial t} = \mathcal{H}(t) |\Psi(t)\rangle, \]
where \(\mathcal{H}(t)\) is the (generally non-Hermitian) Hamiltonian encoding the system and its interaction with the measurement apparatus. The resulting state vector \(|\Psi(t)\rangle\) is not necessarily normalized to unity; its components (i.e., projections onto a basis, such as \(\langle \text{outcome} | \Psi(t) \rangle\)) are interpreted as \textit{relative probability amplitudes}. To obtain physical probabilities for specific outcomes at any given time \(t\), the squared magnitudes of these relative probability amplitudes (yielding relative probabilities) are subsequently normalized by the total instantaneous norm \(\langle\Psi(t)|\Psi(t)\rangle\) (i.e., \(P_{\text{outcome}}(t) = |\langle \text{outcome} | \Psi(t) \rangle|^2 / \langle \Psi(t) | \Psi(t) \rangle\)). We note that alternative theoretical approaches exist where norm preservation is enforced continuously via non-linear modifications to the Schr\"odinger equation (see, e.g., Ref.~\cite{brody_mixed-state_2012}).

    \item \textbf{Closed Loop in Parameter Space:} The Hamiltonian $\mathcal{H}(\mathbf{X}(t))$ traces a closed loop in parameter space $\mathbf{X}(t) = \{X_1(t), X_2(t), ...\}$ during the measurement, ensuring that it returns to its initial form $\mathcal{H}_0$ by the end of the process (Fig. 1(a)). This can be represented as:
    
    \begin{equation}
        \mathcal{H}(\mathbf{X}(t=T)) = \mathcal{H}(\mathbf{X}(t=0)) = \mathcal{H}_0
    \end{equation}
    
    \item \textbf{Non-Hermitian term:} The interaction with the measurement apparatus is modeled by a time-dependent non-Hermitian term in the Hamiltonian. This term also describes a loop in parameter space, potentially dependent on the initial state of the system.
\end{enumerate}

These assumptions ensure that the system returns to its original form after the measurement apparatus is removed. The non-Hermitian term serves as a minimal model for the apparatus's influence, avoiding the need to explicitly include the apparatus's degrees of freedom in the Hilbert space, which would lead to a more complex many-body problem, and is readily implemented in quantum simulators such as photonic lattices~\cite{szameit_pt-symmetry_2011,pan_photonic_2018,miri_exceptional_2019} and electrical circuits~\cite{schindler_experimental_2011,hofmann_reciprocal_2020}. The case of Ref.~\cite{martinez_romeral_wavefunction_2024}, where collapse is obtained in a Stern-Gerlach setup including both spin and position variables does not include an implicit change in time. However, it can be viewed as a particular case where the particle's passage through the magnetic field region effectively introduces the time dependence when viewed on the spin variable.

In essence, our model describes the measurement-induced collapse as resulting from a continuous (and not instantaneous) process, driven by a non-Hermitian perturbation on the initial Hermitian Hamiltonian, which could be mapped to a close loop dynamics in parameter space, dictated by the variables of the Hamiltonian, some of which may be influenced by the initial state.

\section{Possible mechanisms}
Let us now turn to the discussion of the possible mechanisms. There are different ways where a non-Hermitian term could lead to a collapse of a superposition state on one of its components. A very simple way would be just to add gains and losses to amplify a target state and reduce the others. A typical case is shown in Fig.\ref{fig1} (b), where a superposition state is collapsed onto one of its components. In this case, we take for illustration purposes:
\begin{equation}
    {\cal H}={\cal H}_0+\lambda {\cal H}_1.
\label{eq-H}
\end{equation}
where:
\begin{equation}
    {\cal H}_0= \gamma_0
    \begin{pmatrix}
0 & 1\\
1 & 0
\end{pmatrix},
\end{equation}

\begin{equation}
    {\cal H}_1= 
    \begin{pmatrix}
i \Gamma & 0\\
0 & -i \Gamma 
\end{pmatrix}.
\end{equation}
The time dependence is embedded in the parameter $\lambda$ which is taken as a normalized Gaussian centered at the measurement time ($t=5$ in units of $\hbar/ \gamma_0$, represented with a vertical dashed line in Fig.\ref{fig1} (b)) and with a standard deviation  $\sigma=1$. 
Specifically, we take \(\lambda(t) = \exp(-(t-t_{0})^{2}/(2\sigma^{2}))/(\sigma\sqrt{2\pi})\). The state vector \(|\Psi(t)\rangle\) evolves under \(\mathcal{H}\) according to the linear Schr\"odinger equation as outlined in Section II. The probabilities displayed in Fig.1 (b) for having spin up (blue continuous line) and spin down (orange dashed line) as a function of time (the initial state is a spin down state) are then calculated from the components of \(|\Psi(t)\rangle\) -- whose projections onto the spin basis states are treated as relative probability amplitudes -- by squaring their magnitudes and normalizing by the total instantaneous norm \(\langle\Psi(t)|\Psi(t)\rangle\).
The meter acts around $t=5$ amplifying the spin up and reducing the spin down components leading to a situation where the superposition vanishes, the \textit{collapse}. After the interaction with the meter vanishes the usual time-evolution leads again to a superposition state.

Although these models show how non-Hermitian Hamiltonians can amplify a chosen state, the need to construct the interaction specifically to favor a predetermined outcome perhaps offers less insight into the measurement process itself. One might desire a mechanism where state selection emerges more dynamically from the system's response to control or interaction. This motivates our exploration of an alternative framework centered on dynamically induced collapse.

For simplicity, in what follows we will focus on energy measurements. A hint for the search for other possible mechanisms, is that the underlying phenomenon needs to bear an intrinsic asymmetry so that by changing a variable in the ``measurement apparatus" one could obtain the different possible outcomes, it also needs to be a destroyer of superpositions. Furthermore, the mathematical substrate for such phenomenon is likely to break many of the usual rules to serve as a proper backdrop for \textit{the cut}. Within the realm of non-Hermitian systems, one of the most prominent sources of unusual and counter-intuitive physics are the so called exceptional points~\cite{kato_perturbation_1995,heiss_physics_2012}. These are points in parameter space where the eigenvalues and eigenvectors coalesce (see the scheme in Fig.\ref{fig1} (c) illustrating the Riemann surfaces close to an exceptional point), and have been the focus of intense research ~\cite{dembowski_experimental_2001,dembowski_encircling_2004,miri_exceptional_2019}. The coalescence at an exceptional point needs to be distinguished from a usual Hermitian degeneracy where a full set of eigenstates is kept. The key difference being the appearance of \textit{defectiveness}~\cite{foa_torres_perspective_2019}, the lack of a full set of eigenstates. The dependence of the eigenvalues on the parameters around exceptional points is typically non-analytic~\cite{heiss_physics_2012}.

Although for a $2\times 2$ system non-Hermitian system, the adiabatic solutions are expected to exchange when encircling an exceptional point~\cite{heiss_chirality_2001,dembowski_experimental_2001,heiss_physics_2012}, this conversion has been predicted to occur in an asymmetric way~\cite{uzdin_observability_2011,berry_slow_2011}, the so called \textit{chiral state or mode conversion}~\cite{hassan_dynamically_2017} because of non-adiabatic transitions. This phenomenon manifests in a two-state system during the encirclement of an exceptional point within the parameter space. In the realm of Hermitian systems, the adiabatic theorem posits that if a system is initially prepared in one of its eigenstates and the Hamiltonian parameters are gradually altered along a closed trajectory, the system will revert to its initial state, albeit with an additional geometric phase~\cite{berry_quantal_1984}. In contrast, in non-Hermitian systems, a clockwise encirclement around an exceptional point invariably results in the system settling into one of the eigenstates (deemed the preferred state). On the other hand, an anticlockwise journey consistently leads to the emergence of the second eigenstate, as depicted in Fig.\ref{fig1} (d). This intriguing behavior has been observed across various platforms, including 
waveguides~\cite{doppler_dynamically_2016}, and optomechanical systems~\cite{xu_topological_2016}. At a fundamental level, this state conversion has been associated with the Stokes phenomenon of asymptotics~\cite{berry_slow_2011} and the concept of stability loss delay~\cite{milburn_general_2015}.

Remarkably, the dynamics of the instantaneous eigenstates' weights in non-Hermitian systems diverge significantly from those in Hermitian counterparts. The weights can exhibit extreme fluctuations~\cite{berry_slow_2011}, and transitions can occur even under slow dynamics, challenging the conventional understanding of adiabaticity~\cite{doppler_dynamically_2016,berry_slow_2011}. Ref.~\cite{hassan_dynamically_2017} has presented analytical and numerical insights on why a preferred state gets amplified in a dynamical setting. Recent papers indicate that chiral state conversion can occur even without the need to encircle the exceptional point~\cite{hassan_chiral_2017,nasari_observation_2022}, thus broadening the phenomenon's observed scope. But the understanding of the phenomenon of state conversion is still far from complete, as recent studies indicate~\cite{nye_universal_2023,nye_adiabatic_2024,kumar_general_2025}. Clearly, more research is needed to get a better picture of the time scales needed for it to happen and its robustness.

Chiral state conversion provides a promising mechanism for modeling the state vector collapse in a $2\times 2$ system. For this, the interaction with the meter should lead to exceptional points in parameter space. The loop starts (and ends) with the Hamiltonian being ${\cal H}_0$ which is Hermitian. Right at the beginning of the loop, a crucial decision must be made regarding the direction in which the loop will proceed. Switching between clockwise and anticlockwise loops changes the outcome or the preferred state. Upon the loop's completion, when the interaction ceases and the Hamiltonian reverts to its original form ${\cal H}_0$, the initial superpositions that may have been in the initial state disappear, ushering in a state of certainty. This transition marks a pivotal moment in the dynamics of the system, where the ephemeral nature of quantum superpositions gives way to definitive outcomes, encapsulating the essence of quantum measurement and evolution.

To make the concept of forming a closed loop in parameter space more concrete, let us consider a minimal example which captures the essential physics of EP encirclement. A general form for such a Hamiltonian can be written relative to its parameters $\delta(t)$ and $\gamma(t)$ as:
\begin{equation}
    {\cal H}(t) = \begin{pmatrix} \delta(t) + i\gamma(t) & J \\ J & -\delta(t) - i\gamma(t) \end{pmatrix} 
    \label{eq:H_loop_general_revised} 
\end{equation}

Here, $\mathbb{I}$ is the identity matrix and \(J\) represents a real coupling strength, and \(\delta(t)\) is a time-varying detuning. The term \(\gamma(t)\) represents half the time-varying gain/loss difference between the two states. Its significance is central to the non-Hermitian nature of the Hamiltonian during the measurement interaction; it explicitly introduces gain or loss, modeling the strength and character of the system-apparatus coupling that drives the non-unitary evolution. The specific choice and time-dependence of \(\gamma(t)\) (along with \(\delta(t)\)) define the parametric loop in the non-Hermitian parameter space, which is crucial for the proposed chiral state conversion mechanism. The model requires \(\gamma(t)\) to be non-zero only during the transient measurement interaction, returning to zero (as does \(\delta(t)\)) at the beginning (\(t=0\)) and end (\(t=T\)) of the loop, ensuring the Hamiltonian reverts to the Hermitian \(\mathcal{H}_{0}\).
Let the initial and final Hamiltonian be the Hermitian ${\cal H}_0$, corresponding to parameters $\delta=0$, and $\gamma=0$: 
\begin{equation}
    {\cal H}_0 = \begin{pmatrix} 0 & J \\ J & 0 \end{pmatrix}.
    \label{eq:H0_loop_example_revised} 
\end{equation}
The EP occurs at $(\delta_{EP}, \gamma_{EP}) = (0, J)$ (assuming $J>0$). A closed loop in the $(\delta, \gamma)$ parameter space that starts from ${\cal H}_0$ (i.e., starts at $(\delta=0, \gamma=0)$), encircles this EP, and returns to ${\cal H}_0$ at time $t=T$ can be explicitly implemented. For example, consider a circular path centered precisely at the EP $(0, J)$ with radius $R=J$. This path passes through the origin and is parameterized during the interval $0 \le t \le T$ by:
\begin{equation}
    \delta(t) = J \sin(\Omega t)
    \label{eq:delta_loop_revised} 
\end{equation}
and
\begin{equation}
    \gamma(t) = J (1 - \cos(\Omega t)),
    \label{eq:gamma_loop_revised} 
\end{equation}
where $\Omega = 2\pi/T$ determines the loop duration $T$. Note that at $t=0$ and $t=T$, we have $\delta=0$ and $\gamma=0$, ensuring the Hamiltonian starts and ends at ${\cal H}_0$. The time evolution governed by this explicit ${\cal H}(t)$ generates the phenomenon of chiral state 
 conversion~\cite{hassan_dynamically_2017}.

Crucially, the analysis of the non-Hermitian evolution induced by the parameter loop reveals that the eigenstates of the initial Hamiltonian${\cal H}_0$   are the only stable dynamical outcomes. The evolution eliminates components orthogonal to these states, thus providing a dynamical basis for why these are the privileged outcomes. Furthermore, the topology of the path (specifically, the loop direction/chirality) determines which of these preferred states is ultimately selected.

To connect our model to the probabilistic outcomes of quantum measurements, we must address how the choice of loop direction (chirality) relates to Born's rule. Our framework proposes that the non-Hermitian interaction drives the system towards a specific eigenstate depending on the chirality (loop direction). However, the model in its current form does not inherently derive the probabilities associated with these outcomes. To recover the predictions of conventional quantum mechanics, we must impose Born's rule externally: the probability of the interaction selecting a particular chirality (clockwise or anti-clockwise) must be explicitly set to match the squared amplitude of the corresponding component in the system's state vector just before the measurement interaction begins. The precise physical mechanism responsible for enforcing this probabilistic choice based on the initial state remains an open question not solved by this dynamical model. Requiring the chirality selection probability to depend on the initial state amplitudes inherently introduces a non-linear element into the overall measurement process description, distinguishing it from the purely linear Schrödinger evolution between measurements. This aligns with arguments suggesting that a purely linear dynamics cannot fully replicate Born's rule ~\cite{mertens_inconsistency_2021}. Importantly, for a sufficiently slow variation of the parameters, encircling the exceptional point might not be necessary, as the conversion has been signaled to occur anyway~\cite{hassan_chiral_2017,nasari_observation_2022}. This robustness may simplify the experimental implementation of our proposed mechanism.

Looking ahead, a crucial question is how the probabilistic nature of quantum measurement emerges from the seemingly deterministic chiral dynamics. We speculate that, as long as the environmental interactions (at the origin of the non-Hermitian term) are random and unbiased, they will lead to a statistical distribution of chirality choices that mirrors the probabilities encoded in the quantum state, hence statistically aligning with Born’s rule and the predictions of conventional quantum mechanics. Exploring this connection between environmental interactions, non-Hermitian dynamics, and the emergence of Born's rule is a promising direction for future research.

\textit{Collapse when there is no measurement apparatus.--} The collapse of the wave function can occur even in the absence of a dedicated measurement apparatus. While our model focuses on the case where a non-Hermitian term represents the measurement interaction, extending it to other scenarios, such as those involving many-body interactions, remains an open question. A possible path forward could involve exploring the proliferation of exceptional points~\cite{luitz_exceptional_2019} or the phenomena reported in~\cite{znidaric_solvable_2022}. Importantly, any interaction that vanishes after a finite time, leaving the system's Hamiltonian unchanged, automatically satisfies our assumption 2, allowing for the possibility of spontaneous collapse driven by transient interactions represented as loops in parameter space.

\section{Final remarks} 
We have discussed a framework for a dynamical model of a quantum measurement where collapse emerges from an evolution dictated by the Schrödinger equation. A quantum measurement is seen as resulting from the interaction with other degrees of freedom introducing a non-Hermitian term to an otherwise Hermitian Hamiltonian. The nature of this interaction, having a finite lifespan, motivates using a time-dependent Hamiltonian which, throughout the interaction, describes a closed loop in parameter space. As a possible mechanism, we signal the phenomenon called chiral state conversion, which leads to the destruction of superposition states in favor of one of the eigenstates of the initial Hamiltonian (which, for our purposes, is  Hermitian). For a $2\times2$ system, the outcome can be switched by changing the encircling of the loop from clockwise to anticlockwise. 
Crucially, to comply with experimental observations, the probability of the interaction selecting a clockwise versus anti-clockwise loop must be imposed according to Born's rule, determined by the amplitudes of the system's state vector immediately prior to the interaction. As discussed, incorporating this state-dependent probability selection introduces the necessary non-linearity into the description of the measurement process~\cite{mertens_inconsistency_2021}.

Conceptually, our analysis utilizing parameter loops around exceptional points offers insight into two key aspects of quantum measurement. Firstly, it provides a dynamical mechanism for state selection based on path topology (chirality). Secondly, and perhaps more fundamentally, the dynamics inherent to the non-Hermitian evolution during the loop naturally establish the eigenstates of the initial Hamiltonian $\mathcal{H}_0$ as the unique stable outcomes, offering a potential mechanistic explanation for their privileged status in this context. By signaling how the system is dynamically driven towards these states before one is selected, our framework provides a more complete picture of the collapse process.

Chiral mode conversion has been extensively studied for $2\times 2$ systems, but there is still much to understand for systems with more degrees of freedom, where other mechanisms such as the non-Hermitian skin effect~\cite{martinez_alvarez_non-hermitian_2018,yao_edge_2018} (see also ~\cite{bergholtz_exceptional_2021,lin_topological_2023} and references therein) may come handy.

%
As a corollary and perhaps the most immediately useful outcome of our discussion, one can envision the simulation of quantum measurements using classical objects: photonic setups with classical light, electrical circuits, and many others. For the case of photonic waveguide arrays~\cite{szameit_pt-symmetry_2011,vicencio_nonsymmetric_2025}, the direction of propagation $z$ naturally takes the role of time $t$~\cite{rechtsman_photonic_2013}, owing to the direct analogy between the paraxial wave equation for the light envelope and the time-dependent Schr\"odinger equation. The effective Hamiltonian maps onto the structure's properties: coupling $J(z)$ via waveguide spacing, detuning $\delta(z)$ via refractive index $n(x,y,z)$, and non-Hermitian terms $\gamma(z)$ via engineered gain/loss (e.g., complex $n$ or coupling to lossy elements). Since these parameters can be precisely modulated along $z$ during fabrication (e.g., via laser writing), static waveguide circuits can be created to implement specific time-dependent Hamiltonians, including the parameter loops around exceptional points central to our mechanism. This allows for direct observation of the resulting dynamics, including testing the chiral state selection by comparing structures fabricated with opposite loop winding directions along $z$.

We hope that this may trigger a fresh look at this enduring enigma. Perhaps the challenge ahead is to take the measurement out of the quantum realm and down to an experiment with classical objects.

\section{Acknowledgments} 
We thank Joaquín Márquez Delgado, Diego Bautista Avilés, Igor Gornyi, Yuval Gefen, and Mikhail Kiselev for useful discussions, and C. Cormick, Michael Berry, Franck Laloë and Quancheng Liu for useful comments. L.E.F.F.T. acknowledges financial support by ANID FONDECYT (Chile) through grants 1211038 and 1250751, The Abdus Salam International Center for Theoretical Physics and the Simons Foundation, and by the EU Horizon 2020 research and innovation program under the Marie-Sklodowska-Curie Grant Agreement No. 873028 (HYDROTRONICS Project). ICN2 is funded by the CERCA programme / Generalitat de Catalunya, and is supported by the Severo Ochoa Centres of Excellence programme, Grant CEX2021-001214-S, funded by MCIN / AEI / 10.13039.501100011033. This work is also supported by MICIN with European funds‐NextGenerationEU (PRTR‐C17.I1) and by 2021 SGR 00997, funded by Generalitat de Catalunya.

\section{Author Contributions} 
Building on their established research partnership, L. E. F. F. T. and S. R. collaborated on this new line of inquiry starting with a question posed by S. R.. For this paper, L. E. F. F. T. developed the novel concept, led the project, and drafted the manuscript. S. R. contributed valuable insights and critical feedback throughout the process. Both authors were involved in refining the ideas and revising the final manuscript.


\begin{thebibliography}{82}%
\makeatletter
\providecommand \@ifxundefined [1]{%
 \@ifx{#1\undefined}
}%
\providecommand \@ifnum [1]{%
 \ifnum #1\expandafter \@firstoftwo
 \else \expandafter \@secondoftwo
 \fi
}%
\providecommand \@ifx [1]{%
 \ifx #1\expandafter \@firstoftwo
 \else \expandafter \@secondoftwo
 \fi
}%
\providecommand \natexlab [1]{#1}%
\providecommand \enquote  [1]{``#1''}%
\providecommand \bibnamefont  [1]{#1}%
\providecommand \bibfnamefont [1]{#1}%
\providecommand \citenamefont [1]{#1}%
\providecommand \href@noop [0]{\@secondoftwo}%
\providecommand \href [0]{\begingroup \@sanitize@url \@href}%
\providecommand \@href[1]{\@@startlink{#1}\@@href}%
\providecommand \@@href[1]{\endgroup#1\@@endlink}%
\providecommand \@sanitize@url [0]{\catcode `\\12\catcode `\$12\catcode `\&12\catcode `\#12\catcode `\^12\catcode `\_12\catcode `\%12\relax}%
\providecommand \@@startlink[1]{}%
\providecommand \@@endlink[0]{}%
\providecommand \url  [0]{\begingroup\@sanitize@url \@url }%
\providecommand \@url [1]{\endgroup\@href {#1}{\urlprefix }}%
\providecommand \urlprefix  [0]{URL }%
\providecommand \Eprint [0]{\href }%
\providecommand \doibase [0]{http://dx.doi.org/}%
\providecommand \selectlanguage [0]{\@gobble}%
\providecommand \bibinfo  [0]{\@secondoftwo}%
\providecommand \bibfield  [0]{\@secondoftwo}%
\providecommand \translation [1]{[#1]}%
\providecommand \BibitemOpen [0]{}%
\providecommand \bibitemStop [0]{}%
\providecommand \bibitemNoStop [0]{.\EOS\space}%
\providecommand \EOS [0]{\spacefactor3000\relax}%
\providecommand \BibitemShut  [1]{\csname bibitem#1\endcsname}%
\let\auto@bib@innerbib\@empty
\bibitem [{\citenamefont {Bell}(1990)}]{bell_against_1990}%
  \BibitemOpen
  \bibfield  {author} {\bibinfo {author} {\bibfnamefont {J.}~\bibnamefont {Bell}},\ }\bibfield  {title} {{\selectlanguage {en}\enquote {\bibinfo {title} {Against `measurement'},}\ }}\href {\doibase 10.1088/2058-7058/3/8/26} {\bibfield  {journal} {\bibinfo  {journal} {Physics World}\ }\textbf {\bibinfo {volume} {3}},\ \bibinfo {pages} {33} (\bibinfo {year} {1990})}\BibitemShut {NoStop}%
\bibitem [{\citenamefont {Dirac}(1963)}]{dirac_evolution_1963}%
  \BibitemOpen
  \bibfield  {author} {\bibinfo {author} {\bibfnamefont {P.~A.~M.}\ \bibnamefont {Dirac}},\ }\bibfield  {title} {\enquote {\bibinfo {title} {The {Evolution} of the {Physicist}’s {Picture} of {Nature}},}\ }\href {https://www.jstor.org/stable/24936146} {\bibfield  {journal} {\bibinfo  {journal} {Scientific American}\ }\textbf {\bibinfo {volume} {208}},\ \bibinfo {pages} {45} (\bibinfo {year} {1963})}\BibitemShut {NoStop}%
\bibitem [{\citenamefont {Laloë}(2001)}]{laloe_we_2001}%
  \BibitemOpen
  \bibfield  {author} {\bibinfo {author} {\bibfnamefont {F.}~\bibnamefont {Laloë}},\ }\bibfield  {title} {\enquote {\bibinfo {title} {Do we really understand quantum mechanics? {Strange} correlations, paradoxes, and theorems},}\ }\href {\doibase 10.1119/1.1356698} {\bibfield  {journal} {\bibinfo  {journal} {American Journal of Physics}\ }\textbf {\bibinfo {volume} {69}},\ \bibinfo {pages} {655} (\bibinfo {year} {2001})}\BibitemShut {NoStop}%
\bibitem [{\citenamefont {Cohen-Tannoudji}\ \emph {et~al.}(2020)\citenamefont {Cohen-Tannoudji}, \citenamefont {Diu},\ and\ \citenamefont {Laloë}}]{cohen-tannoudji_quantum_2020}%
  \BibitemOpen
  \bibfield  {author} {\bibinfo {author} {\bibfnamefont {C.}~\bibnamefont {Cohen-Tannoudji}}, \bibinfo {author} {\bibfnamefont {B.}~\bibnamefont {Diu}}, \ and\ \bibinfo {author} {\bibfnamefont {F.}~\bibnamefont {Laloë}},\ }\href {https://www.wiley.com/en-us/Quantum+Mechanics%2C+Volume+1%3A+Basic+Concepts%2C+Tools%2C+and+Applications%2C+2nd+Edition-p-9783527822713} {{\selectlanguage {en-us}\emph {\bibinfo {title} {Quantum {Mechanics}, {Volume} 1: {Basic} {Concepts}, {Tools}, and {Applications}, 2nd {Edition}}}}}\ (\bibinfo  {publisher} {Wiley},\ \bibinfo {year} {2020})\BibitemShut {NoStop}%
\bibitem [{\citenamefont {Shankar}(1994)}]{shankar_principles_1994}%
  \BibitemOpen
  \bibfield  {author} {\bibinfo {author} {\bibfnamefont {R.}~\bibnamefont {Shankar}},\ }\href {\doibase 10.1007/978-1-4757-0576-8} {{\selectlanguage {en}\emph {\bibinfo {title} {Principles of {Quantum} {Mechanics}}}}},\ \bibinfo {edition} {2nd}\ ed.\ (\bibinfo  {publisher} {Springer US},\ \bibinfo {year} {1994})\BibitemShut {NoStop}%
\bibitem [{\citenamefont {Bacciagaluppi}\ and\ \citenamefont {Crull}(2011)}]{bacciagaluppi_translation_2011}%
  \BibitemOpen
  \bibfield  {author} {\bibinfo {author} {\bibfnamefont {G.}~\bibnamefont {Bacciagaluppi}}\ and\ \bibinfo {author} {\bibfnamefont {E.}~\bibnamefont {Crull}},\ }\bibfield  {title} {{\selectlanguage {English}\enquote {\bibinfo {title} {Translation of: {W}. {Heisenberg}, {Ist} eine deterministische {Ergänzung} der {Quantenmechanik} möglich?}}\ }}\href {https://abdn.pure.elsevier.com/en/publications/translation-of-w-heisenberg-ist-eine-deterministische-erg%C3%A4nzung-d} {\  (\bibinfo {year} {2011})}\BibitemShut {NoStop}%
\bibitem [{\citenamefont {Einstein}\ \emph {et~al.}(1935)\citenamefont {Einstein}, \citenamefont {Podolsky},\ and\ \citenamefont {Rosen}}]{einstein_can_1935}%
  \BibitemOpen
  \bibfield  {author} {\bibinfo {author} {\bibfnamefont {A.}~\bibnamefont {Einstein}}, \bibinfo {author} {\bibfnamefont {B.}~\bibnamefont {Podolsky}}, \ and\ \bibinfo {author} {\bibfnamefont {N.}~\bibnamefont {Rosen}},\ }\bibfield  {title} {\enquote {\bibinfo {title} {Can {Quantum}-{Mechanical} {Description} of {Physical} {Reality} {Be} {Considered} {Complete}?}}\ }\href {\doibase 10.1103/PhysRev.47.777} {\bibfield  {journal} {\bibinfo  {journal} {Physical Review}\ }\textbf {\bibinfo {volume} {47}},\ \bibinfo {pages} {777} (\bibinfo {year} {1935})}\BibitemShut {NoStop}%
\bibitem [{\citenamefont {Pearle}(2000)}]{pearle_wavefunction_2000}%
  \BibitemOpen
  \bibfield  {author} {\bibinfo {author} {\bibfnamefont {P.}~\bibnamefont {Pearle}},\ }\bibfield  {title} {{\selectlanguage {en}\enquote {\bibinfo {title} {Wavefunction {Collapse} and {Conservation} {Laws}},}\ }}\href {\doibase 10.1023/A:1003677103804} {\bibfield  {journal} {\bibinfo  {journal} {Foundations of Physics}\ }\textbf {\bibinfo {volume} {30}},\ \bibinfo {pages} {1145} (\bibinfo {year} {2000})}\BibitemShut {NoStop}%
\bibitem [{\citenamefont {Wheeler}(2016)}]{wheeler_law_2016}%
  \BibitemOpen
  \bibfield  {author} {\bibinfo {author} {\bibfnamefont {J.~A.}\ \bibnamefont {Wheeler}},\ }\bibfield  {title} {{\selectlanguage {en}\enquote {\bibinfo {title} {Law {Without} {Law}},}\ }}in\ \href {https://press.princeton.edu/books/hardcover/9780691641027/quantum-theory-and-measurement} {{\selectlanguage {en}\emph {\bibinfo {booktitle} {Quantum {Theory} and {Measurement}, edited by {John} {Archibald} {Wheeler} and {Wojciech} {Hubert} {Zurek}}}}}\ (\bibinfo  {publisher} {Princeton University Press},\ \bibinfo {year} {2016})\BibitemShut {NoStop}%
\bibitem [{Note1()}]{Note1}%
  \BibitemOpen
  \bibinfo {note} {Wheeler mentions that the term \protect \textit {irreversible amplification} was used by Bohr, see chapter 1 of~\cite {bohr_atomic_2010}, which was published in 1958.}\BibitemShut {Stop}%
\bibitem [{\citenamefont {Zurek}(2003)}]{zurek_decoherence_2003}%
  \BibitemOpen
  \bibfield  {author} {\bibinfo {author} {\bibfnamefont {W.~H.}\ \bibnamefont {Zurek}},\ }\bibfield  {title} {\enquote {\bibinfo {title} {Decoherence, einselection, and the quantum origins of the classical},}\ }\href {\doibase 10.1103/RevModPhys.75.715} {\bibfield  {journal} {\bibinfo  {journal} {Reviews of Modern Physics}\ }\textbf {\bibinfo {volume} {75}},\ \bibinfo {pages} {715} (\bibinfo {year} {2003})}\BibitemShut {NoStop}%
\bibitem [{\citenamefont {Schlosshauer}(2007)}]{schlosshauer_m_decoherence_2007}%
  \BibitemOpen
  \bibfield  {author} {\bibinfo {author} {\bibfnamefont {M.}~\bibnamefont {Schlosshauer}},\ }\href {\doibase 10.1007/978-3-540-35775-9} {{\selectlanguage {en}\emph {\bibinfo {title} {Decoherence and the {Quantum}-{To}-{Classical} {Transition}}}}},\ Frontiers {Collection}\ (\bibinfo  {publisher} {Springer},\ \bibinfo {address} {Berlin, Heidelberg},\ \bibinfo {year} {2007})\BibitemShut {NoStop}%
\bibitem [{\citenamefont {Leggett}(2006)}]{leggett_problems_2006}%
  \BibitemOpen
  \bibfield  {author} {\bibinfo {author} {\bibfnamefont {A.}~\bibnamefont {Leggett}},\ }\href@noop {} {\emph {\bibinfo {title} {The {Problems} of {Physics}}}},\ Oxford {Classic} {Texts} in the {Physical} {Sciences}\ (\bibinfo  {publisher} {Oxford University Press},\ \bibinfo {address} {Oxford, New York},\ \bibinfo {year} {2006})\BibitemShut {NoStop}%
\bibitem [{\citenamefont {Hance}\ and\ \citenamefont {Hossenfelder}(2022)}]{hance_what_2022}%
  \BibitemOpen
  \bibfield  {author} {\bibinfo {author} {\bibfnamefont {J.~R.}\ \bibnamefont {Hance}}\ and\ \bibinfo {author} {\bibfnamefont {S.}~\bibnamefont {Hossenfelder}},\ }\bibfield  {title} {{\selectlanguage {en}\enquote {\bibinfo {title} {What does it take to solve the measurement problem?}}\ }}\href {\doibase 10.1088/2399-6528/ac96cf} {\bibfield  {journal} {\bibinfo  {journal} {Journal of Physics Communications}\ }\textbf {\bibinfo {volume} {6}},\ \bibinfo {pages} {102001} (\bibinfo {year} {2022})}\BibitemShut {NoStop}%
\bibitem [{Note2()}]{Note2}%
  \BibitemOpen
  \bibinfo {note} {There are also notable opinions against thinking of the measurement problem as such, see for example~\cite {mermin_there_2022}. We also refer to~\cite {schlosshauer_measurement_2011} for a broader overview of the different positions}\BibitemShut {NoStop}%
\bibitem [{\citenamefont {Ghirardi}\ \emph {et~al.}(1986)\citenamefont {Ghirardi}, \citenamefont {Rimini},\ and\ \citenamefont {Weber}}]{ghirardi_unified_1986}%
  \BibitemOpen
  \bibfield  {author} {\bibinfo {author} {\bibfnamefont {G.~C.}\ \bibnamefont {Ghirardi}}, \bibinfo {author} {\bibfnamefont {A.}~\bibnamefont {Rimini}}, \ and\ \bibinfo {author} {\bibfnamefont {T.}~\bibnamefont {Weber}},\ }\bibfield  {title} {\enquote {\bibinfo {title} {Unified dynamics for microscopic and macroscopic systems},}\ }\href {\doibase 10.1103/PhysRevD.34.470} {\bibfield  {journal} {\bibinfo  {journal} {Physical Review D}\ }\textbf {\bibinfo {volume} {34}},\ \bibinfo {pages} {470} (\bibinfo {year} {1986})}\BibitemShut {NoStop}%
\bibitem [{\citenamefont {Ghirardi}\ \emph {et~al.}(1990)\citenamefont {Ghirardi}, \citenamefont {Pearle},\ and\ \citenamefont {Rimini}}]{ghirardi_markov_1990}%
  \BibitemOpen
  \bibfield  {author} {\bibinfo {author} {\bibfnamefont {G.~C.}\ \bibnamefont {Ghirardi}}, \bibinfo {author} {\bibfnamefont {P.}~\bibnamefont {Pearle}}, \ and\ \bibinfo {author} {\bibfnamefont {A.}~\bibnamefont {Rimini}},\ }\bibfield  {title} {\enquote {\bibinfo {title} {Markov processes in {Hilbert} space and continuous spontaneous localization of systems of identical particles},}\ }\href {\doibase 10.1103/PhysRevA.42.78} {\bibfield  {journal} {\bibinfo  {journal} {Physical Review A}\ }\textbf {\bibinfo {volume} {42}},\ \bibinfo {pages} {78} (\bibinfo {year} {1990})}\BibitemShut {NoStop}%
\bibitem [{\citenamefont {Pearle}(1989)}]{pearle_combining_1989}%
  \BibitemOpen
  \bibfield  {author} {\bibinfo {author} {\bibfnamefont {P.}~\bibnamefont {Pearle}},\ }\bibfield  {title} {\enquote {\bibinfo {title} {Combining stochastic dynamical state-vector reduction with spontaneous localization},}\ }\href {\doibase 10.1103/PhysRevA.39.2277} {\bibfield  {journal} {\bibinfo  {journal} {Physical Review A}\ }\textbf {\bibinfo {volume} {39}},\ \bibinfo {pages} {2277} (\bibinfo {year} {1989})}\BibitemShut {NoStop}%
\bibitem [{\citenamefont {Bassi}\ and\ \citenamefont {Ghirardi}(2003)}]{bassi_dynamical_2003}%
  \BibitemOpen
  \bibfield  {author} {\bibinfo {author} {\bibfnamefont {A.}~\bibnamefont {Bassi}}\ and\ \bibinfo {author} {\bibfnamefont {G.}~\bibnamefont {Ghirardi}},\ }\bibfield  {title} {\enquote {\bibinfo {title} {Dynamical reduction models},}\ }\href {\doibase 10.1016/S0370-1573(03)00103-0} {\bibfield  {journal} {\bibinfo  {journal} {Physics Reports}\ }\textbf {\bibinfo {volume} {379}},\ \bibinfo {pages} {257} (\bibinfo {year} {2003})}\BibitemShut {NoStop}%
\bibitem [{\citenamefont {Fernández-Alcázar}\ and\ \citenamefont {Pastawski}(2015)}]{fernandez-alcazar_decoherent_2015}%
  \BibitemOpen
  \bibfield  {author} {\bibinfo {author} {\bibfnamefont {L.~J.}\ \bibnamefont {Fernández-Alcázar}}\ and\ \bibinfo {author} {\bibfnamefont {H.~M.}\ \bibnamefont {Pastawski}},\ }\bibfield  {title} {\enquote {\bibinfo {title} {Decoherent time-dependent transport beyond the {Landauer}-{B}{\textbackslash}"uttiker formulation: {A} quantum-drift alternative to quantum jumps},}\ }\href {\doibase 10.1103/PhysRevA.91.022117} {\bibfield  {journal} {\bibinfo  {journal} {Physical Review A}\ }\textbf {\bibinfo {volume} {91}},\ \bibinfo {pages} {022117} (\bibinfo {year} {2015})}\BibitemShut {NoStop}%
\bibitem [{\citenamefont {Diósi}(1989)}]{diosi_models_1989}%
  \BibitemOpen
  \bibfield  {author} {\bibinfo {author} {\bibfnamefont {L.}~\bibnamefont {Diósi}},\ }\bibfield  {title} {\enquote {\bibinfo {title} {Models for universal reduction of macroscopic quantum fluctuations},}\ }\href {\doibase 10.1103/PhysRevA.40.1165} {\bibfield  {journal} {\bibinfo  {journal} {Physical Review A}\ }\textbf {\bibinfo {volume} {40}},\ \bibinfo {pages} {1165} (\bibinfo {year} {1989})}\BibitemShut {NoStop}%
\bibitem [{\citenamefont {Donadi}\ \emph {et~al.}(2021)\citenamefont {Donadi}, \citenamefont {Piscicchia}, \citenamefont {Curceanu}, \citenamefont {Diósi}, \citenamefont {Laubenstein},\ and\ \citenamefont {Bassi}}]{donadi_underground_2021}%
  \BibitemOpen
  \bibfield  {author} {\bibinfo {author} {\bibfnamefont {S.}~\bibnamefont {Donadi}}, \bibinfo {author} {\bibfnamefont {K.}~\bibnamefont {Piscicchia}}, \bibinfo {author} {\bibfnamefont {C.}~\bibnamefont {Curceanu}}, \bibinfo {author} {\bibfnamefont {L.}~\bibnamefont {Diósi}}, \bibinfo {author} {\bibfnamefont {M.}~\bibnamefont {Laubenstein}}, \ and\ \bibinfo {author} {\bibfnamefont {A.}~\bibnamefont {Bassi}},\ }\bibfield  {title} {{\selectlanguage {en}\enquote {\bibinfo {title} {Underground test of gravity-related wave function collapse},}\ }}\href {\doibase 10.1038/s41567-020-1008-4} {\bibfield  {journal} {\bibinfo  {journal} {Nature Physics}\ }\textbf {\bibinfo {volume} {17}},\ \bibinfo {pages} {74} (\bibinfo {year} {2021})}\BibitemShut {NoStop}%
\bibitem [{\citenamefont {Penrose}(1996)}]{penrose_gravitys_1996}%
  \BibitemOpen
  \bibfield  {author} {\bibinfo {author} {\bibfnamefont {R.}~\bibnamefont {Penrose}},\ }\bibfield  {title} {{\selectlanguage {en}\enquote {\bibinfo {title} {On {Gravity}'s role in {Quantum} {State} {Reduction}},}\ }}\href {\doibase 10.1007/BF02105068} {\bibfield  {journal} {\bibinfo  {journal} {General Relativity and Gravitation}\ }\textbf {\bibinfo {volume} {28}},\ \bibinfo {pages} {581} (\bibinfo {year} {1996})}\BibitemShut {NoStop}%
\bibitem [{\citenamefont {Martínez~Romeral}\ \emph {et~al.}(2024)\citenamefont {Martínez~Romeral}, \citenamefont {Foa~Torres},\ and\ \citenamefont {Roche}}]{martinez_romeral_wavefunction_2024}%
  \BibitemOpen
  \bibfield  {author} {\bibinfo {author} {\bibfnamefont {J.}~\bibnamefont {Martínez~Romeral}}, \bibinfo {author} {\bibfnamefont {L.~E.~F.}\ \bibnamefont {Foa~Torres}}, \ and\ \bibinfo {author} {\bibfnamefont {S.}~\bibnamefont {Roche}},\ }\bibfield  {title} {{\selectlanguage {en}\enquote {\bibinfo {title} {Wavefunction collapse driven by non-{Hermitian} disturbance},}\ }}\href {\doibase 10.1088/2399-6528/ad5b37} {\bibfield  {journal} {\bibinfo  {journal} {Journal of Physics Communications}\ }\textbf {\bibinfo {volume} {8}},\ \bibinfo {pages} {071001} (\bibinfo {year} {2024})}\BibitemShut {NoStop}%
\bibitem [{\citenamefont {Singh}\ \emph {et~al.}(2023{\natexlab{a}})\citenamefont {Singh}, \citenamefont {Singh},\ and\ \citenamefont {Banerjee}}]{singh_embedding_2023}%
  \BibitemOpen
  \bibfield  {author} {\bibinfo {author} {\bibfnamefont {G.}~\bibnamefont {Singh}}, \bibinfo {author} {\bibfnamefont {R.~K.}\ \bibnamefont {Singh}}, \ and\ \bibinfo {author} {\bibfnamefont {S.}~\bibnamefont {Banerjee}},\ }\bibfield  {title} {{\selectlanguage {en}\enquote {\bibinfo {title} {Embedding of a non-{Hermitian} {Hamiltonian} to emulate the von {Neumann} measurement scheme},}\ }}\href {\doibase 10.1088/1751-8121/ad1431} {\bibfield  {journal} {\bibinfo  {journal} {Journal of Physics A: Mathematical and Theoretical}\ }\textbf {\bibinfo {volume} {57}},\ \bibinfo {pages} {035301} (\bibinfo {year} {2023}{\natexlab{a}})}\BibitemShut {NoStop}%
\bibitem [{\citenamefont {Singh}\ \emph {et~al.}(2023{\natexlab{b}})\citenamefont {Singh}, \citenamefont {Singh},\ and\ \citenamefont {Banerjee}}]{singh_emulating_2023}%
  \BibitemOpen
  \bibfield  {author} {\bibinfo {author} {\bibfnamefont {G.}~\bibnamefont {Singh}}, \bibinfo {author} {\bibfnamefont {R.~K.}\ \bibnamefont {Singh}}, \ and\ \bibinfo {author} {\bibfnamefont {S.}~\bibnamefont {Banerjee}},\ }\href {\doibase 10.48550/arXiv.2302.01898} {\enquote {\bibinfo {title} {Emulating the measurement postulates of quantum mechanics via non-{Hermitian} {Hamiltonian}},}\ } (\bibinfo {year} {2023}{\natexlab{b}}),\ \bibinfo {note} {arXiv:2302.01898 [quant-ph]}\BibitemShut {NoStop}%
\bibitem [{\citenamefont {Van~Wezel}(2010)}]{van_wezel_broken_2010}%
  \BibitemOpen
  \bibfield  {author} {\bibinfo {author} {\bibfnamefont {J.}~\bibnamefont {Van~Wezel}},\ }\bibfield  {title} {{\selectlanguage {en}\enquote {\bibinfo {title} {Broken {Time} {Translation} {Symmetry} as a {Model} for {Quantum} {State} {Reduction}},}\ }}\href {\doibase 10.3390/sym2020582} {\bibfield  {journal} {\bibinfo  {journal} {Symmetry}\ }\textbf {\bibinfo {volume} {2}},\ \bibinfo {pages} {582} (\bibinfo {year} {2010})}\BibitemShut {NoStop}%
\bibitem [{\citenamefont {Mertens}\ \emph {et~al.}(2023)\citenamefont {Mertens}, \citenamefont {Wesseling},\ and\ \citenamefont {van Wezel}}]{mertens_objective_2023}%
  \BibitemOpen
  \bibfield  {author} {\bibinfo {author} {\bibfnamefont {L.}~\bibnamefont {Mertens}}, \bibinfo {author} {\bibfnamefont {M.}~\bibnamefont {Wesseling}}, \ and\ \bibinfo {author} {\bibfnamefont {J.}~\bibnamefont {van Wezel}},\ }\bibfield  {title} {{\selectlanguage {en}\enquote {\bibinfo {title} {An objective collapse model without state dependent stochasticity},}\ }}\href {\doibase 10.21468/SciPostPhys.14.5.114} {\bibfield  {journal} {\bibinfo  {journal} {SciPost Physics}\ }\textbf {\bibinfo {volume} {14}},\ \bibinfo {pages} {114} (\bibinfo {year} {2023})}\BibitemShut {NoStop}%
\bibitem [{\citenamefont {Mukherjee}\ and\ \citenamefont {van Wezel}(2024)}]{mukherjee_colored-noise-driven_2024}%
  \BibitemOpen
  \bibfield  {author} {\bibinfo {author} {\bibfnamefont {A.}~\bibnamefont {Mukherjee}}\ and\ \bibinfo {author} {\bibfnamefont {J.}~\bibnamefont {van Wezel}},\ }\bibfield  {title} {\enquote {\bibinfo {title} {Colored-noise-driven unitarity violation causing dynamical quantum state reduction},}\ }\href {\doibase 10.1103/PhysRevA.109.032214} {\bibfield  {journal} {\bibinfo  {journal} {Physical Review A}\ }\textbf {\bibinfo {volume} {109}},\ \bibinfo {pages} {032214} (\bibinfo {year} {2024})}\BibitemShut {NoStop}%
\bibitem [{\citenamefont {Gisin}\ and\ \citenamefont {Piron}(1981)}]{gisin_collapse_1981}%
  \BibitemOpen
  \bibfield  {author} {\bibinfo {author} {\bibfnamefont {N.}~\bibnamefont {Gisin}}\ and\ \bibinfo {author} {\bibfnamefont {C.}~\bibnamefont {Piron}},\ }\bibfield  {title} {{\selectlanguage {En}\enquote {\bibinfo {title} {Collapse of the wave packet without mixture},}\ }}\href {\doibase 10.1007/BF02285309} {\bibfield  {journal} {\bibinfo  {journal} {Letters in Mathematical Physics}\ }\textbf {\bibinfo {volume} {5}},\ \bibinfo {pages} {379} (\bibinfo {year} {1981})}\BibitemShut {NoStop}%
\bibitem [{\citenamefont {Bassi}\ \emph {et~al.}(2013)\citenamefont {Bassi}, \citenamefont {Lochan}, \citenamefont {Satin}, \citenamefont {Singh},\ and\ \citenamefont {Ulbricht}}]{bassi_models_2013}%
  \BibitemOpen
  \bibfield  {author} {\bibinfo {author} {\bibfnamefont {A.}~\bibnamefont {Bassi}}, \bibinfo {author} {\bibfnamefont {K.}~\bibnamefont {Lochan}}, \bibinfo {author} {\bibfnamefont {S.}~\bibnamefont {Satin}}, \bibinfo {author} {\bibfnamefont {T.~P.}\ \bibnamefont {Singh}}, \ and\ \bibinfo {author} {\bibfnamefont {H.}~\bibnamefont {Ulbricht}},\ }\bibfield  {title} {\enquote {\bibinfo {title} {Models of wave-function collapse, underlying theories, and experimental tests},}\ }\href {\doibase 10.1103/RevModPhys.85.471} {\bibfield  {journal} {\bibinfo  {journal} {Reviews of Modern Physics}\ }\textbf {\bibinfo {volume} {85}},\ \bibinfo {pages} {471} (\bibinfo {year} {2013})}\BibitemShut {NoStop}%
\bibitem [{\citenamefont {Rechtsman}\ \emph {et~al.}(2013)\citenamefont {Rechtsman}, \citenamefont {Zeuner}, \citenamefont {Plotnik}, \citenamefont {Lumer}, \citenamefont {Podolsky}, \citenamefont {Dreisow}, \citenamefont {Nolte}, \citenamefont {Segev},\ and\ \citenamefont {Szameit}}]{rechtsman_photonic_2013}%
  \BibitemOpen
  \bibfield  {author} {\bibinfo {author} {\bibfnamefont {M.~C.}\ \bibnamefont {Rechtsman}}, \bibinfo {author} {\bibfnamefont {J.~M.}\ \bibnamefont {Zeuner}}, \bibinfo {author} {\bibfnamefont {Y.}~\bibnamefont {Plotnik}}, \bibinfo {author} {\bibfnamefont {Y.}~\bibnamefont {Lumer}}, \bibinfo {author} {\bibfnamefont {D.}~\bibnamefont {Podolsky}}, \bibinfo {author} {\bibfnamefont {F.}~\bibnamefont {Dreisow}}, \bibinfo {author} {\bibfnamefont {S.}~\bibnamefont {Nolte}}, \bibinfo {author} {\bibfnamefont {M.}~\bibnamefont {Segev}}, \ and\ \bibinfo {author} {\bibfnamefont {A.}~\bibnamefont {Szameit}},\ }\bibfield  {title} {\enquote {\bibinfo {title} {Photonic {Floquet} topological insulators},}\ }\href {http://dx.doi.org/10.1038/nature12066} {\bibfield  {journal} {\bibinfo  {journal} {Nature}\ }\textbf {\bibinfo {volume} {496}},\ \bibinfo {pages} {196} (\bibinfo {year} {2013})}\BibitemShut {NoStop}%
\bibitem [{\citenamefont {Vicencio}\ \emph {et~al.}(2015)\citenamefont {Vicencio}, \citenamefont {Cantillano}, \citenamefont {Morales-Inostroza}, \citenamefont {Real}, \citenamefont {Mejía-Cortés}, \citenamefont {Weimann}, \citenamefont {Szameit},\ and\ \citenamefont {Molina}}]{vicencio_observation_2015}%
  \BibitemOpen
  \bibfield  {author} {\bibinfo {author} {\bibfnamefont {R.~A.}\ \bibnamefont {Vicencio}}, \bibinfo {author} {\bibfnamefont {C.}~\bibnamefont {Cantillano}}, \bibinfo {author} {\bibfnamefont {L.}~\bibnamefont {Morales-Inostroza}}, \bibinfo {author} {\bibfnamefont {B.}~\bibnamefont {Real}}, \bibinfo {author} {\bibfnamefont {C.}~\bibnamefont {Mejía-Cortés}}, \bibinfo {author} {\bibfnamefont {S.}~\bibnamefont {Weimann}}, \bibinfo {author} {\bibfnamefont {A.}~\bibnamefont {Szameit}}, \ and\ \bibinfo {author} {\bibfnamefont {M.~I.}\ \bibnamefont {Molina}},\ }\bibfield  {title} {\enquote {\bibinfo {title} {Observation of {Localized} {States} in {Lieb} {Photonic} {Lattices}},}\ }\href {\doibase 10.1103/PhysRevLett.114.245503} {\bibfield  {journal} {\bibinfo  {journal} {Physical Review Letters}\ }\textbf {\bibinfo {volume} {114}},\ \bibinfo {pages} {245503} (\bibinfo {year} {2015})}\BibitemShut {NoStop}%
\bibitem [{\citenamefont {Veenstra}\ \emph {et~al.}(2024)\citenamefont {Veenstra}, \citenamefont {Gamayun}, \citenamefont {Guo}, \citenamefont {Sarvi}, \citenamefont {Meinersen},\ and\ \citenamefont {Coulais}}]{veenstra_non-reciprocal_2024}%
  \BibitemOpen
  \bibfield  {author} {\bibinfo {author} {\bibfnamefont {J.}~\bibnamefont {Veenstra}}, \bibinfo {author} {\bibfnamefont {O.}~\bibnamefont {Gamayun}}, \bibinfo {author} {\bibfnamefont {X.}~\bibnamefont {Guo}}, \bibinfo {author} {\bibfnamefont {A.}~\bibnamefont {Sarvi}}, \bibinfo {author} {\bibfnamefont {C.~V.}\ \bibnamefont {Meinersen}}, \ and\ \bibinfo {author} {\bibfnamefont {C.}~\bibnamefont {Coulais}},\ }\bibfield  {title} {{\selectlanguage {en}\enquote {\bibinfo {title} {Non-reciprocal topological solitons in active metamaterials},}\ }}\href {\doibase 10.1038/s41586-024-07097-6} {\bibfield  {journal} {\bibinfo  {journal} {Nature}\ }\textbf {\bibinfo {volume} {627}},\ \bibinfo {pages} {528} (\bibinfo {year} {2024})}\BibitemShut {NoStop}%
\bibitem [{\citenamefont {Zhang}\ \emph {et~al.}(2021)\citenamefont {Zhang}, \citenamefont {Yang}, \citenamefont {Ge}, \citenamefont {Guan}, \citenamefont {Chen}, \citenamefont {Yan}, \citenamefont {Chen}, \citenamefont {Xi}, \citenamefont {Li}, \citenamefont {Jia}, \citenamefont {Yuan}, \citenamefont {Sun}, \citenamefont {Chen},\ and\ \citenamefont {Zhang}}]{zhang_acoustic_2021}%
  \BibitemOpen
  \bibfield  {author} {\bibinfo {author} {\bibfnamefont {L.}~\bibnamefont {Zhang}}, \bibinfo {author} {\bibfnamefont {Y.}~\bibnamefont {Yang}}, \bibinfo {author} {\bibfnamefont {Y.}~\bibnamefont {Ge}}, \bibinfo {author} {\bibfnamefont {Y.-J.}\ \bibnamefont {Guan}}, \bibinfo {author} {\bibfnamefont {Q.}~\bibnamefont {Chen}}, \bibinfo {author} {\bibfnamefont {Q.}~\bibnamefont {Yan}}, \bibinfo {author} {\bibfnamefont {F.}~\bibnamefont {Chen}}, \bibinfo {author} {\bibfnamefont {R.}~\bibnamefont {Xi}}, \bibinfo {author} {\bibfnamefont {Y.}~\bibnamefont {Li}}, \bibinfo {author} {\bibfnamefont {D.}~\bibnamefont {Jia}}, \bibinfo {author} {\bibfnamefont {S.-Q.}\ \bibnamefont {Yuan}}, \bibinfo {author} {\bibfnamefont {H.-X.}\ \bibnamefont {Sun}}, \bibinfo {author} {\bibfnamefont {H.}~\bibnamefont {Chen}}, \ and\ \bibinfo {author} {\bibfnamefont {B.}~\bibnamefont {Zhang}},\ }\bibfield  {title} {{\selectlanguage {en}\enquote {\bibinfo {title} {Acoustic non-{Hermitian} skin effect from twisted winding topology},}\ }}\href
  {\doibase 10.1038/s41467-021-26619-8} {\bibfield  {journal} {\bibinfo  {journal} {Nature Communications}\ }\textbf {\bibinfo {volume} {12}},\ \bibinfo {pages} {6297} (\bibinfo {year} {2021})}\BibitemShut {NoStop}%
\bibitem [{\citenamefont {Lorenz}\ \emph {et~al.}(2023)\citenamefont {Lorenz}, \citenamefont {Kohler}, \citenamefont {Parafilo}, \citenamefont {Kiselev},\ and\ \citenamefont {Ludwig}}]{lorenz_classical_2023}%
  \BibitemOpen
  \bibfield  {author} {\bibinfo {author} {\bibfnamefont {H.}~\bibnamefont {Lorenz}}, \bibinfo {author} {\bibfnamefont {S.}~\bibnamefont {Kohler}}, \bibinfo {author} {\bibfnamefont {A.}~\bibnamefont {Parafilo}}, \bibinfo {author} {\bibfnamefont {M.}~\bibnamefont {Kiselev}}, \ and\ \bibinfo {author} {\bibfnamefont {S.}~\bibnamefont {Ludwig}},\ }\bibfield  {title} {{\selectlanguage {en}\enquote {\bibinfo {title} {Classical analogue to driven quantum bits based on macroscopic pendula},}\ }}\href {\doibase 10.1038/s41598-023-45118-y} {\bibfield  {journal} {\bibinfo  {journal} {Scientific Reports}\ }\textbf {\bibinfo {volume} {13}},\ \bibinfo {pages} {18386} (\bibinfo {year} {2023})}\BibitemShut {NoStop}%
\bibitem [{\citenamefont {Lee}\ \emph {et~al.}(2018)\citenamefont {Lee}, \citenamefont {Imhof}, \citenamefont {Berger}, \citenamefont {Bayer}, \citenamefont {Brehm}, \citenamefont {Molenkamp}, \citenamefont {Kiessling},\ and\ \citenamefont {Thomale}}]{lee_topolectrical_2018}%
  \BibitemOpen
  \bibfield  {author} {\bibinfo {author} {\bibfnamefont {C.~H.}\ \bibnamefont {Lee}}, \bibinfo {author} {\bibfnamefont {S.}~\bibnamefont {Imhof}}, \bibinfo {author} {\bibfnamefont {C.}~\bibnamefont {Berger}}, \bibinfo {author} {\bibfnamefont {F.}~\bibnamefont {Bayer}}, \bibinfo {author} {\bibfnamefont {J.}~\bibnamefont {Brehm}}, \bibinfo {author} {\bibfnamefont {L.~W.}\ \bibnamefont {Molenkamp}}, \bibinfo {author} {\bibfnamefont {T.}~\bibnamefont {Kiessling}}, \ and\ \bibinfo {author} {\bibfnamefont {R.}~\bibnamefont {Thomale}},\ }\bibfield  {title} {{\selectlanguage {en}\enquote {\bibinfo {title} {Topolectrical {Circuits}},}\ }}\href {\doibase 10.1038/s42005-018-0035-2} {\bibfield  {journal} {\bibinfo  {journal} {Communications Physics}\ }\textbf {\bibinfo {volume} {1}},\ \bibinfo {pages} {1} (\bibinfo {year} {2018})}\BibitemShut {NoStop}%
\bibitem [{\citenamefont {Szameit}\ and\ \citenamefont {Nolte}(2010)}]{szameit_discrete_2010}%
  \BibitemOpen
  \bibfield  {author} {\bibinfo {author} {\bibfnamefont {A.}~\bibnamefont {Szameit}}\ and\ \bibinfo {author} {\bibfnamefont {S.}~\bibnamefont {Nolte}},\ }\bibfield  {title} {{\selectlanguage {en}\enquote {\bibinfo {title} {Discrete optics in femtosecond-laser-written photonic structures},}\ }}\href {\doibase 10.1088/0953-4075/43/16/163001} {\bibfield  {journal} {\bibinfo  {journal} {Journal of Physics B: Atomic, Molecular and Optical Physics}\ }\textbf {\bibinfo {volume} {43}},\ \bibinfo {pages} {163001} (\bibinfo {year} {2010})}\BibitemShut {NoStop}%
\bibitem [{\citenamefont {Ezawa}(2019)}]{ezawa_electric-circuit_2019}%
  \BibitemOpen
  \bibfield  {author} {\bibinfo {author} {\bibfnamefont {M.}~\bibnamefont {Ezawa}},\ }\bibfield  {title} {\enquote {\bibinfo {title} {Electric-circuit simulation of the {Schrödinger} equation and non-{Hermitian} quantum walks},}\ }\href {\doibase 10.1103/PhysRevB.100.165419} {\bibfield  {journal} {\bibinfo  {journal} {Physical Review B}\ }\textbf {\bibinfo {volume} {100}},\ \bibinfo {pages} {165419} (\bibinfo {year} {2019})}\BibitemShut {NoStop}%
\bibitem [{\citenamefont {Longhi}(2006)}]{longhi_nonexponential_2006}%
  \BibitemOpen
  \bibfield  {author} {\bibinfo {author} {\bibfnamefont {S.}~\bibnamefont {Longhi}},\ }\bibfield  {title} {\enquote {\bibinfo {title} {Nonexponential {Decay} {Via} {Tunneling} in {Tight}-{Binding} {Lattices} and the {Optical} {Zeno} {Effect}},}\ }\href {\doibase 10.1103/PhysRevLett.97.110402} {\bibfield  {journal} {\bibinfo  {journal} {Physical Review Letters}\ }\textbf {\bibinfo {volume} {97}} (\bibinfo {year} {2006}),\ 10.1103/PhysRevLett.97.110402}\BibitemShut {NoStop}%
\bibitem [{\citenamefont {Liu}\ \emph {et~al.}(2023)\citenamefont {Liu}, \citenamefont {Liu}, \citenamefont {Ziegler},\ and\ \citenamefont {Chen}}]{liu_engineering_2023}%
  \BibitemOpen
  \bibfield  {author} {\bibinfo {author} {\bibfnamefont {Q.}~\bibnamefont {Liu}}, \bibinfo {author} {\bibfnamefont {W.}~\bibnamefont {Liu}}, \bibinfo {author} {\bibfnamefont {K.}~\bibnamefont {Ziegler}}, \ and\ \bibinfo {author} {\bibfnamefont {F.}~\bibnamefont {Chen}},\ }\bibfield  {title} {\enquote {\bibinfo {title} {Engineering of {Zeno} {Dynamics} in {Integrated} {Photonics}},}\ }\href {\doibase 10.1103/PhysRevLett.130.103801} {\bibfield  {journal} {\bibinfo  {journal} {Physical Review Letters}\ }\textbf {\bibinfo {volume} {130}},\ \bibinfo {pages} {103801} (\bibinfo {year} {2023})}\BibitemShut {NoStop}%
\bibitem [{\citenamefont {Moiseyev}(2011)}]{moiseyev_non-hermitian_2011}%
  \BibitemOpen
  \bibfield  {author} {\bibinfo {author} {\bibfnamefont {N.}~\bibnamefont {Moiseyev}},\ }\href {http://www.cambridge.org/us/academic/subjects/physics/quantum-physics-quantum-information-and-quantum-computation/non-hermitian-quantum-mechanics} {\emph {\bibinfo {title} {Non-{Hermitian} {Quantum} {Mechanics}}}}\ (\bibinfo  {publisher} {Cambridge University Press},\ \bibinfo {year} {2011})\BibitemShut {NoStop}%
\bibitem [{\citenamefont {Bergholtz}\ \emph {et~al.}(2021)\citenamefont {Bergholtz}, \citenamefont {Budich},\ and\ \citenamefont {Kunst}}]{bergholtz_exceptional_2021}%
  \BibitemOpen
  \bibfield  {author} {\bibinfo {author} {\bibfnamefont {E.~J.}\ \bibnamefont {Bergholtz}}, \bibinfo {author} {\bibfnamefont {J.~C.}\ \bibnamefont {Budich}}, \ and\ \bibinfo {author} {\bibfnamefont {F.~K.}\ \bibnamefont {Kunst}},\ }\bibfield  {title} {\enquote {\bibinfo {title} {Exceptional topology of non-{Hermitian} systems},}\ }\href {\doibase 10.1103/RevModPhys.93.015005} {\bibfield  {journal} {\bibinfo  {journal} {Reviews of Modern Physics}\ }\textbf {\bibinfo {volume} {93}},\ \bibinfo {pages} {015005} (\bibinfo {year} {2021})}\BibitemShut {NoStop}%
\bibitem [{\citenamefont {Lin}\ \emph {et~al.}(2023)\citenamefont {Lin}, \citenamefont {Tai}, \citenamefont {Li},\ and\ \citenamefont {Lee}}]{lin_topological_2023}%
  \BibitemOpen
  \bibfield  {author} {\bibinfo {author} {\bibfnamefont {R.}~\bibnamefont {Lin}}, \bibinfo {author} {\bibfnamefont {T.}~\bibnamefont {Tai}}, \bibinfo {author} {\bibfnamefont {L.}~\bibnamefont {Li}}, \ and\ \bibinfo {author} {\bibfnamefont {C.~H.}\ \bibnamefont {Lee}},\ }\bibfield  {title} {{\selectlanguage {en}\enquote {\bibinfo {title} {Topological non-{Hermitian} skin effect},}\ }}\href {\doibase 10.1007/s11467-023-1309-z} {\bibfield  {journal} {\bibinfo  {journal} {Frontiers of Physics}\ }\textbf {\bibinfo {volume} {18}},\ \bibinfo {pages} {53605} (\bibinfo {year} {2023})}\BibitemShut {NoStop}%
\bibitem [{\citenamefont {Kato}(1995)}]{kato_perturbation_1995}%
  \BibitemOpen
  \bibfield  {author} {\bibinfo {author} {\bibfnamefont {T.}~\bibnamefont {Kato}},\ }\href {https://www.springer.com/us/book/9783540586616} {{\selectlanguage {en}\emph {\bibinfo {title} {Perturbation {Theory} for {Linear} {Operators}}}}},\ \bibinfo {edition} {2nd}\ ed.,\ Classics in {Mathematics}\ (\bibinfo  {publisher} {Springer-Verlag},\ \bibinfo {address} {Berlin Heidelberg},\ \bibinfo {year} {1995})\BibitemShut {NoStop}%
\bibitem [{\citenamefont {Heiss}\ and\ \citenamefont {Harney}(2001)}]{heiss_chirality_2001}%
  \BibitemOpen
  \bibfield  {author} {\bibinfo {author} {\bibfnamefont {W.~D.}\ \bibnamefont {Heiss}}\ and\ \bibinfo {author} {\bibfnamefont {H.~L.}\ \bibnamefont {Harney}},\ }\bibfield  {title} {{\selectlanguage {en}\enquote {\bibinfo {title} {The chirality of exceptional points},}\ }}\href {\doibase 10.1007/s100530170017} {\bibfield  {journal} {\bibinfo  {journal} {The European Physical Journal D - Atomic, Molecular, Optical and Plasma Physics}\ }\textbf {\bibinfo {volume} {17}},\ \bibinfo {pages} {149} (\bibinfo {year} {2001})}\BibitemShut {NoStop}%
\bibitem [{\citenamefont {Dembowski}\ \emph {et~al.}(2001)\citenamefont {Dembowski}, \citenamefont {Gräf}, \citenamefont {Harney}, \citenamefont {Heine}, \citenamefont {Heiss}, \citenamefont {Rehfeld},\ and\ \citenamefont {Richter}}]{dembowski_experimental_2001}%
  \BibitemOpen
  \bibfield  {author} {\bibinfo {author} {\bibfnamefont {C.}~\bibnamefont {Dembowski}}, \bibinfo {author} {\bibfnamefont {H.-D.}\ \bibnamefont {Gräf}}, \bibinfo {author} {\bibfnamefont {H.~L.}\ \bibnamefont {Harney}}, \bibinfo {author} {\bibfnamefont {A.}~\bibnamefont {Heine}}, \bibinfo {author} {\bibfnamefont {W.~D.}\ \bibnamefont {Heiss}}, \bibinfo {author} {\bibfnamefont {H.}~\bibnamefont {Rehfeld}}, \ and\ \bibinfo {author} {\bibfnamefont {A.}~\bibnamefont {Richter}},\ }\bibfield  {title} {\enquote {\bibinfo {title} {Experimental {Observation} of the {Topological} {Structure} of {Exceptional} {Points}},}\ }\href {\doibase 10.1103/PhysRevLett.86.787} {\bibfield  {journal} {\bibinfo  {journal} {Physical Review Letters}\ }\textbf {\bibinfo {volume} {86}},\ \bibinfo {pages} {787} (\bibinfo {year} {2001})}\BibitemShut {NoStop}%
\bibitem [{\citenamefont {Heiss}(2012)}]{heiss_physics_2012}%
  \BibitemOpen
  \bibfield  {author} {\bibinfo {author} {\bibfnamefont {W.~D.}\ \bibnamefont {Heiss}},\ }\bibfield  {title} {{\selectlanguage {en}\enquote {\bibinfo {title} {The physics of exceptional points},}\ }}\href {\doibase 10.1088/1751-8113/45/44/444016} {\bibfield  {journal} {\bibinfo  {journal} {Journal of Physics A: Mathematical and Theoretical}\ }\textbf {\bibinfo {volume} {45}},\ \bibinfo {pages} {444016} (\bibinfo {year} {2012})}\BibitemShut {NoStop}%
\bibitem [{\citenamefont {Wiersig}(2014)}]{wiersig_enhancing_2014}%
  \BibitemOpen
  \bibfield  {author} {\bibinfo {author} {\bibfnamefont {J.}~\bibnamefont {Wiersig}},\ }\bibfield  {title} {\enquote {\bibinfo {title} {Enhancing the {Sensitivity} of {Frequency} and {Energy} {Splitting} {Detection} by {Using} {Exceptional} {Points}: {Application} to {Microcavity} {Sensors} for {Single}-{Particle} {Detection}},}\ }\href {\doibase 10.1103/PhysRevLett.112.203901} {\bibfield  {journal} {\bibinfo  {journal} {Physical Review Letters}\ }\textbf {\bibinfo {volume} {112}},\ \bibinfo {pages} {203901} (\bibinfo {year} {2014})}\BibitemShut {NoStop}%
\bibitem [{\citenamefont {Rechtsman}(2017)}]{rechtsman_optical_2017}%
  \BibitemOpen
  \bibfield  {author} {\bibinfo {author} {\bibfnamefont {M.~C.}\ \bibnamefont {Rechtsman}},\ }\bibfield  {title} {{\selectlanguage {en}\enquote {\bibinfo {title} {Optical sensing gets exceptional},}\ }}\href {\doibase 10.1038/548161a} {\bibfield  {journal} {\bibinfo  {journal} {Nature}\ }\textbf {\bibinfo {volume} {548}},\ \bibinfo {pages} {161} (\bibinfo {year} {2017})}\BibitemShut {NoStop}%
\bibitem [{\citenamefont {Hodaei}\ \emph {et~al.}(2017)\citenamefont {Hodaei}, \citenamefont {Hassan}, \citenamefont {Wittek}, \citenamefont {Garcia-Gracia}, \citenamefont {El-Ganainy}, \citenamefont {Christodoulides},\ and\ \citenamefont {Khajavikhan}}]{hodaei_enhanced_2017}%
  \BibitemOpen
  \bibfield  {author} {\bibinfo {author} {\bibfnamefont {H.}~\bibnamefont {Hodaei}}, \bibinfo {author} {\bibfnamefont {A.~U.}\ \bibnamefont {Hassan}}, \bibinfo {author} {\bibfnamefont {S.}~\bibnamefont {Wittek}}, \bibinfo {author} {\bibfnamefont {H.}~\bibnamefont {Garcia-Gracia}}, \bibinfo {author} {\bibfnamefont {R.}~\bibnamefont {El-Ganainy}}, \bibinfo {author} {\bibfnamefont {D.~N.}\ \bibnamefont {Christodoulides}}, \ and\ \bibinfo {author} {\bibfnamefont {M.}~\bibnamefont {Khajavikhan}},\ }\bibfield  {title} {{\selectlanguage {en}\enquote {\bibinfo {title} {Enhanced sensitivity at higher-order exceptional points},}\ }}\href {\doibase 10.1038/nature23280} {\bibfield  {journal} {\bibinfo  {journal} {Nature}\ }\textbf {\bibinfo {volume} {548}},\ \bibinfo {pages} {187} (\bibinfo {year} {2017})}\BibitemShut {NoStop}%
\bibitem [{\citenamefont {Chen}\ \emph {et~al.}(2017)\citenamefont {Chen}, \citenamefont {Kaya~Özdemir}, \citenamefont {Zhao}, \citenamefont {Wiersig},\ and\ \citenamefont {Yang}}]{chen_exceptional_2017}%
  \BibitemOpen
  \bibfield  {author} {\bibinfo {author} {\bibfnamefont {W.}~\bibnamefont {Chen}}, \bibinfo {author} {\bibfnamefont {{\c S}.}~\bibnamefont {Kaya~Özdemir}}, \bibinfo {author} {\bibfnamefont {G.}~\bibnamefont {Zhao}}, \bibinfo {author} {\bibfnamefont {J.}~\bibnamefont {Wiersig}}, \ and\ \bibinfo {author} {\bibfnamefont {L.}~\bibnamefont {Yang}},\ }\bibfield  {title} {{\selectlanguage {en}\enquote {\bibinfo {title} {Exceptional points enhance sensing in an optical microcavity},}\ }}\href {\doibase 10.1038/nature23281} {\bibfield  {journal} {\bibinfo  {journal} {Nature}\ }\textbf {\bibinfo {volume} {548}},\ \bibinfo {pages} {192} (\bibinfo {year} {2017})}\BibitemShut {NoStop}%
\bibitem [{\citenamefont {Hokmabadi}\ \emph {et~al.}(2019)\citenamefont {Hokmabadi}, \citenamefont {Schumer}, \citenamefont {Christodoulides},\ and\ \citenamefont {Khajavikhan}}]{hokmabadi_non-hermitian_2019}%
  \BibitemOpen
  \bibfield  {author} {\bibinfo {author} {\bibfnamefont {M.~P.}\ \bibnamefont {Hokmabadi}}, \bibinfo {author} {\bibfnamefont {A.}~\bibnamefont {Schumer}}, \bibinfo {author} {\bibfnamefont {D.~N.}\ \bibnamefont {Christodoulides}}, \ and\ \bibinfo {author} {\bibfnamefont {M.}~\bibnamefont {Khajavikhan}},\ }\bibfield  {title} {{\selectlanguage {en}\enquote {\bibinfo {title} {Non-{Hermitian} ring laser gyroscopes with enhanced {Sagnac} sensitivity},}\ }}\href {\doibase 10.1038/s41586-019-1780-4} {\bibfield  {journal} {\bibinfo  {journal} {Nature}\ }\textbf {\bibinfo {volume} {576}},\ \bibinfo {pages} {70} (\bibinfo {year} {2019})}\BibitemShut {NoStop}%
\bibitem [{\citenamefont {Hassan}\ \emph {et~al.}(2017{\natexlab{a}})\citenamefont {Hassan}, \citenamefont {Zhen}, \citenamefont {Soljačić}, \citenamefont {Khajavikhan},\ and\ \citenamefont {Christodoulides}}]{hassan_dynamically_2017}%
  \BibitemOpen
  \bibfield  {author} {\bibinfo {author} {\bibfnamefont {A.~U.}\ \bibnamefont {Hassan}}, \bibinfo {author} {\bibfnamefont {B.}~\bibnamefont {Zhen}}, \bibinfo {author} {\bibfnamefont {M.}~\bibnamefont {Soljačić}}, \bibinfo {author} {\bibfnamefont {M.}~\bibnamefont {Khajavikhan}}, \ and\ \bibinfo {author} {\bibfnamefont {D.~N.}\ \bibnamefont {Christodoulides}},\ }\bibfield  {title} {\enquote {\bibinfo {title} {Dynamically {Encircling} {Exceptional} {Points}: {Exact} {Evolution} and {Polarization} {State} {Conversion}},}\ }\href {\doibase 10.1103/PhysRevLett.118.093002} {\bibfield  {journal} {\bibinfo  {journal} {Physical Review Letters}\ }\textbf {\bibinfo {volume} {118}},\ \bibinfo {pages} {093002} (\bibinfo {year} {2017}{\natexlab{a}})}\BibitemShut {NoStop}%
\bibitem [{\citenamefont {Szameit}\ \emph {et~al.}(2011)\citenamefont {Szameit}, \citenamefont {Rechtsman}, \citenamefont {Bahat-Treidel},\ and\ \citenamefont {Segev}}]{szameit_pt-symmetry_2011}%
  \BibitemOpen
  \bibfield  {author} {\bibinfo {author} {\bibfnamefont {A.}~\bibnamefont {Szameit}}, \bibinfo {author} {\bibfnamefont {M.~C.}\ \bibnamefont {Rechtsman}}, \bibinfo {author} {\bibfnamefont {O.}~\bibnamefont {Bahat-Treidel}}, \ and\ \bibinfo {author} {\bibfnamefont {M.}~\bibnamefont {Segev}},\ }\bibfield  {title} {\enquote {\bibinfo {title} {{PT}-symmetry in honeycomb photonic lattices},}\ }\href {\doibase 10.1103/PhysRevA.84.021806} {\bibfield  {journal} {\bibinfo  {journal} {Physical Review A}\ }\textbf {\bibinfo {volume} {84}},\ \bibinfo {pages} {021806} (\bibinfo {year} {2011})}\BibitemShut {NoStop}%
\bibitem [{\citenamefont {Pan}\ \emph {et~al.}(2018)\citenamefont {Pan}, \citenamefont {Zhao}, \citenamefont {Miao}, \citenamefont {Longhi},\ and\ \citenamefont {Feng}}]{pan_photonic_2018}%
  \BibitemOpen
  \bibfield  {author} {\bibinfo {author} {\bibfnamefont {M.}~\bibnamefont {Pan}}, \bibinfo {author} {\bibfnamefont {H.}~\bibnamefont {Zhao}}, \bibinfo {author} {\bibfnamefont {P.}~\bibnamefont {Miao}}, \bibinfo {author} {\bibfnamefont {S.}~\bibnamefont {Longhi}}, \ and\ \bibinfo {author} {\bibfnamefont {L.}~\bibnamefont {Feng}},\ }\bibfield  {title} {{\selectlanguage {en}\enquote {\bibinfo {title} {Photonic zero mode in a non-{Hermitian} photonic lattice},}\ }}\href {\doibase 10.1038/s41467-018-03822-8} {\bibfield  {journal} {\bibinfo  {journal} {Nature Communications}\ }\textbf {\bibinfo {volume} {9}},\ \bibinfo {pages} {1308} (\bibinfo {year} {2018})}\BibitemShut {NoStop}%
\bibitem [{\citenamefont {Miri}\ and\ \citenamefont {Alù}(2019)}]{miri_exceptional_2019}%
  \BibitemOpen
  \bibfield  {author} {\bibinfo {author} {\bibfnamefont {M.-A.}\ \bibnamefont {Miri}}\ and\ \bibinfo {author} {\bibfnamefont {A.}~\bibnamefont {Alù}},\ }\bibfield  {title} {{\selectlanguage {en}\enquote {\bibinfo {title} {Exceptional points in optics and photonics},}\ }}\href {\doibase 10.1126/science.aar7709} {\bibfield  {journal} {\bibinfo  {journal} {Science}\ }\textbf {\bibinfo {volume} {363}},\ \bibinfo {pages} {eaar7709} (\bibinfo {year} {2019})}\BibitemShut {NoStop}%
\bibitem [{\citenamefont {Vicencio}\ \emph {et~al.}(2025)\citenamefont {Vicencio}, \citenamefont {Román-Cortés}, \citenamefont {Rubio-Saldías}, \citenamefont {Vildoso},\ and\ \citenamefont {Foa~Torres}}]{vicencio_nonsymmetric_2025}%
  \BibitemOpen
  \bibfield  {author} {\bibinfo {author} {\bibfnamefont {R.~A.}\ \bibnamefont {Vicencio}}, \bibinfo {author} {\bibfnamefont {D.}~\bibnamefont {Román-Cortés}}, \bibinfo {author} {\bibfnamefont {M.}~\bibnamefont {Rubio-Saldías}}, \bibinfo {author} {\bibfnamefont {P.}~\bibnamefont {Vildoso}}, \ and\ \bibinfo {author} {\bibfnamefont {L.~E.~F.}\ \bibnamefont {Foa~Torres}},\ }\bibfield  {title} {\enquote {\bibinfo {title} {Nonsymmetric evanescent coupling in photonics},}\ }\href {\doibase 10.1103/PhysRevA.111.043510} {\bibfield  {journal} {\bibinfo  {journal} {Physical Review A}\ }\textbf {\bibinfo {volume} {111}},\ \bibinfo {pages} {043510} (\bibinfo {year} {2025})}\BibitemShut {NoStop}%
\bibitem [{\citenamefont {Schindler}\ \emph {et~al.}(2011)\citenamefont {Schindler}, \citenamefont {Li}, \citenamefont {Zheng}, \citenamefont {Ellis},\ and\ \citenamefont {Kottos}}]{schindler_experimental_2011}%
  \BibitemOpen
  \bibfield  {author} {\bibinfo {author} {\bibfnamefont {J.}~\bibnamefont {Schindler}}, \bibinfo {author} {\bibfnamefont {A.}~\bibnamefont {Li}}, \bibinfo {author} {\bibfnamefont {M.~C.}\ \bibnamefont {Zheng}}, \bibinfo {author} {\bibfnamefont {F.~M.}\ \bibnamefont {Ellis}}, \ and\ \bibinfo {author} {\bibfnamefont {T.}~\bibnamefont {Kottos}},\ }\bibfield  {title} {\enquote {\bibinfo {title} {Experimental study of active {LRC} circuits with {PT} symmetries},}\ }\href {\doibase 10.1103/PhysRevA.84.040101} {\bibfield  {journal} {\bibinfo  {journal} {Physical Review A}\ }\textbf {\bibinfo {volume} {84}},\ \bibinfo {pages} {040101} (\bibinfo {year} {2011})}\BibitemShut {NoStop}%
\bibitem [{\citenamefont {Hofmann}\ \emph {et~al.}(2020)\citenamefont {Hofmann}, \citenamefont {Helbig}, \citenamefont {Schindler}, \citenamefont {Salgo}, \citenamefont {Brzezińska}, \citenamefont {Greiter}, \citenamefont {Kiessling}, \citenamefont {Wolf}, \citenamefont {Vollhardt}, \citenamefont {Kabaši}, \citenamefont {Lee}, \citenamefont {Bilušić}, \citenamefont {Thomale},\ and\ \citenamefont {Neupert}}]{hofmann_reciprocal_2020}%
  \BibitemOpen
  \bibfield  {author} {\bibinfo {author} {\bibfnamefont {T.}~\bibnamefont {Hofmann}}, \bibinfo {author} {\bibfnamefont {T.}~\bibnamefont {Helbig}}, \bibinfo {author} {\bibfnamefont {F.}~\bibnamefont {Schindler}}, \bibinfo {author} {\bibfnamefont {N.}~\bibnamefont {Salgo}}, \bibinfo {author} {\bibfnamefont {M.}~\bibnamefont {Brzezińska}}, \bibinfo {author} {\bibfnamefont {M.}~\bibnamefont {Greiter}}, \bibinfo {author} {\bibfnamefont {T.}~\bibnamefont {Kiessling}}, \bibinfo {author} {\bibfnamefont {D.}~\bibnamefont {Wolf}}, \bibinfo {author} {\bibfnamefont {A.}~\bibnamefont {Vollhardt}}, \bibinfo {author} {\bibfnamefont {A.}~\bibnamefont {Kabaši}}, \bibinfo {author} {\bibfnamefont {C.~H.}\ \bibnamefont {Lee}}, \bibinfo {author} {\bibfnamefont {A.}~\bibnamefont {Bilušić}}, \bibinfo {author} {\bibfnamefont {R.}~\bibnamefont {Thomale}}, \ and\ \bibinfo {author} {\bibfnamefont {T.}~\bibnamefont {Neupert}},\ }\bibfield  {title} {\enquote {\bibinfo {title} {Reciprocal skin effect and its realization in a
  topolectrical circuit},}\ }\href {\doibase 10.1103/PhysRevResearch.2.023265} {\bibfield  {journal} {\bibinfo  {journal} {Physical Review Research}\ }\textbf {\bibinfo {volume} {2}},\ \bibinfo {pages} {023265} (\bibinfo {year} {2020})}\BibitemShut {NoStop}%
\bibitem [{\citenamefont {Brody}\ and\ \citenamefont {Graefe}(2012)}]{brody_mixed-state_2012}%
  \BibitemOpen
  \bibfield  {author} {\bibinfo {author} {\bibfnamefont {D.~C.}\ \bibnamefont {Brody}}\ and\ \bibinfo {author} {\bibfnamefont {E.-M.}\ \bibnamefont {Graefe}},\ }\bibfield  {title} {\enquote {\bibinfo {title} {Mixed-{State} {Evolution} in the {Presence} of {Gain} and {Loss}},}\ }\href {\doibase 10.1103/PhysRevLett.109.230405} {\bibfield  {journal} {\bibinfo  {journal} {Physical Review Letters}\ }\textbf {\bibinfo {volume} {109}},\ \bibinfo {pages} {230405} (\bibinfo {year} {2012})}\BibitemShut {NoStop}%
\bibitem [{\citenamefont {Dembowski}\ \emph {et~al.}(2004)\citenamefont {Dembowski}, \citenamefont {Dietz}, \citenamefont {Gräf}, \citenamefont {Harney}, \citenamefont {Heine}, \citenamefont {Heiss},\ and\ \citenamefont {Richter}}]{dembowski_encircling_2004}%
  \BibitemOpen
  \bibfield  {author} {\bibinfo {author} {\bibfnamefont {C.}~\bibnamefont {Dembowski}}, \bibinfo {author} {\bibfnamefont {B.}~\bibnamefont {Dietz}}, \bibinfo {author} {\bibfnamefont {H.-D.}\ \bibnamefont {Gräf}}, \bibinfo {author} {\bibfnamefont {H.~L.}\ \bibnamefont {Harney}}, \bibinfo {author} {\bibfnamefont {A.}~\bibnamefont {Heine}}, \bibinfo {author} {\bibfnamefont {W.~D.}\ \bibnamefont {Heiss}}, \ and\ \bibinfo {author} {\bibfnamefont {A.}~\bibnamefont {Richter}},\ }\bibfield  {title} {\enquote {\bibinfo {title} {Encircling an exceptional point},}\ }\href {\doibase 10.1103/PhysRevE.69.056216} {\bibfield  {journal} {\bibinfo  {journal} {Physical Review E}\ }\textbf {\bibinfo {volume} {69}},\ \bibinfo {pages} {056216} (\bibinfo {year} {2004})}\BibitemShut {NoStop}%
\bibitem [{\citenamefont {Foa~Torres}(2019)}]{foa_torres_perspective_2019}%
  \BibitemOpen
  \bibfield  {author} {\bibinfo {author} {\bibfnamefont {L.~E.~F.}\ \bibnamefont {Foa~Torres}},\ }\bibfield  {title} {{\selectlanguage {en}\enquote {\bibinfo {title} {Perspective on topological states of non-{Hermitian} lattices},}\ }}\href {\doibase 10.1088/2515-7639/ab4092} {\bibfield  {journal} {\bibinfo  {journal} {Journal of Physics: Materials}\ }\textbf {\bibinfo {volume} {3}},\ \bibinfo {pages} {014002} (\bibinfo {year} {2019})}\BibitemShut {NoStop}%
\bibitem [{\citenamefont {Uzdin}\ \emph {et~al.}(2011)\citenamefont {Uzdin}, \citenamefont {Mailybaev},\ and\ \citenamefont {Moiseyev}}]{uzdin_observability_2011}%
  \BibitemOpen
  \bibfield  {author} {\bibinfo {author} {\bibfnamefont {R.}~\bibnamefont {Uzdin}}, \bibinfo {author} {\bibfnamefont {A.}~\bibnamefont {Mailybaev}}, \ and\ \bibinfo {author} {\bibfnamefont {N.}~\bibnamefont {Moiseyev}},\ }\bibfield  {title} {{\selectlanguage {en}\enquote {\bibinfo {title} {On the observability and asymmetry of adiabatic state flips generated by exceptional points},}\ }}\href {\doibase 10.1088/1751-8113/44/43/435302} {\bibfield  {journal} {\bibinfo  {journal} {Journal of Physics A: Mathematical and Theoretical}\ }\textbf {\bibinfo {volume} {44}},\ \bibinfo {pages} {435302} (\bibinfo {year} {2011})}\BibitemShut {NoStop}%
\bibitem [{\citenamefont {Berry}\ and\ \citenamefont {Uzdin}(2011)}]{berry_slow_2011}%
  \BibitemOpen
  \bibfield  {author} {\bibinfo {author} {\bibfnamefont {M.~V.}\ \bibnamefont {Berry}}\ and\ \bibinfo {author} {\bibfnamefont {R.}~\bibnamefont {Uzdin}},\ }\bibfield  {title} {{\selectlanguage {en}\enquote {\bibinfo {title} {Slow non-{Hermitian} cycling: exact solutions and the {Stokes} phenomenon},}\ }}\href {\doibase 10.1088/1751-8113/44/43/435303} {\bibfield  {journal} {\bibinfo  {journal} {Journal of Physics A: Mathematical and Theoretical}\ }\textbf {\bibinfo {volume} {44}},\ \bibinfo {pages} {435303} (\bibinfo {year} {2011})}\BibitemShut {NoStop}%
\bibitem [{\citenamefont {Berry}(1984)}]{berry_quantal_1984}%
  \BibitemOpen
  \bibfield  {author} {\bibinfo {author} {\bibfnamefont {M.~V.}\ \bibnamefont {Berry}},\ }\bibfield  {title} {\enquote {\bibinfo {title} {Quantal {Phase} {Factors} {Accompanying} {Adiabatic} {Changes}},}\ }\href {https://www.jstor.org/stable/2397741} {\bibfield  {journal} {\bibinfo  {journal} {Proceedings of the Royal Society of London. Series A, Mathematical and Physical Sciences}\ }\textbf {\bibinfo {volume} {392}},\ \bibinfo {pages} {45} (\bibinfo {year} {1984})}\BibitemShut {NoStop}%
\bibitem [{\citenamefont {Doppler}\ \emph {et~al.}(2016)\citenamefont {Doppler}, \citenamefont {Mailybaev}, \citenamefont {Böhm}, \citenamefont {Kuhl}, \citenamefont {Girschik}, \citenamefont {Libisch}, \citenamefont {Milburn}, \citenamefont {Rabl}, \citenamefont {Moiseyev},\ and\ \citenamefont {Rotter}}]{doppler_dynamically_2016}%
  \BibitemOpen
  \bibfield  {author} {\bibinfo {author} {\bibfnamefont {J.}~\bibnamefont {Doppler}}, \bibinfo {author} {\bibfnamefont {A.~A.}\ \bibnamefont {Mailybaev}}, \bibinfo {author} {\bibfnamefont {J.}~\bibnamefont {Böhm}}, \bibinfo {author} {\bibfnamefont {U.}~\bibnamefont {Kuhl}}, \bibinfo {author} {\bibfnamefont {A.}~\bibnamefont {Girschik}}, \bibinfo {author} {\bibfnamefont {F.}~\bibnamefont {Libisch}}, \bibinfo {author} {\bibfnamefont {T.~J.}\ \bibnamefont {Milburn}}, \bibinfo {author} {\bibfnamefont {P.}~\bibnamefont {Rabl}}, \bibinfo {author} {\bibfnamefont {N.}~\bibnamefont {Moiseyev}}, \ and\ \bibinfo {author} {\bibfnamefont {S.}~\bibnamefont {Rotter}},\ }\bibfield  {title} {{\selectlanguage {en}\enquote {\bibinfo {title} {Dynamically encircling an exceptional point for asymmetric mode switching},}\ }}\href {\doibase 10.1038/nature18605} {\bibfield  {journal} {\bibinfo  {journal} {Nature}\ }\textbf {\bibinfo {volume} {537}},\ \bibinfo {pages} {76} (\bibinfo {year} {2016})}\BibitemShut {NoStop}%
\bibitem [{\citenamefont {Xu}\ \emph {et~al.}(2016)\citenamefont {Xu}, \citenamefont {Mason}, \citenamefont {Jiang},\ and\ \citenamefont {Harris}}]{xu_topological_2016}%
  \BibitemOpen
  \bibfield  {author} {\bibinfo {author} {\bibfnamefont {H.}~\bibnamefont {Xu}}, \bibinfo {author} {\bibfnamefont {D.}~\bibnamefont {Mason}}, \bibinfo {author} {\bibfnamefont {L.}~\bibnamefont {Jiang}}, \ and\ \bibinfo {author} {\bibfnamefont {J.~G.~E.}\ \bibnamefont {Harris}},\ }\bibfield  {title} {{\selectlanguage {en}\enquote {\bibinfo {title} {Topological energy transfer in an optomechanical system with exceptional points},}\ }}\href {\doibase 10.1038/nature18604} {\bibfield  {journal} {\bibinfo  {journal} {Nature}\ }\textbf {\bibinfo {volume} {537}},\ \bibinfo {pages} {80} (\bibinfo {year} {2016})}\BibitemShut {NoStop}%
\bibitem [{\citenamefont {Milburn}\ \emph {et~al.}(2015)\citenamefont {Milburn}, \citenamefont {Doppler}, \citenamefont {Holmes}, \citenamefont {Portolan}, \citenamefont {Rotter},\ and\ \citenamefont {Rabl}}]{milburn_general_2015}%
  \BibitemOpen
  \bibfield  {author} {\bibinfo {author} {\bibfnamefont {T.~J.}\ \bibnamefont {Milburn}}, \bibinfo {author} {\bibfnamefont {J.}~\bibnamefont {Doppler}}, \bibinfo {author} {\bibfnamefont {C.~A.}\ \bibnamefont {Holmes}}, \bibinfo {author} {\bibfnamefont {S.}~\bibnamefont {Portolan}}, \bibinfo {author} {\bibfnamefont {S.}~\bibnamefont {Rotter}}, \ and\ \bibinfo {author} {\bibfnamefont {P.}~\bibnamefont {Rabl}},\ }\bibfield  {title} {\enquote {\bibinfo {title} {General description of quasiadiabatic dynamical phenomena near exceptional points},}\ }\href {\doibase 10.1103/PhysRevA.92.052124} {\bibfield  {journal} {\bibinfo  {journal} {Physical Review A}\ }\textbf {\bibinfo {volume} {92}},\ \bibinfo {pages} {052124} (\bibinfo {year} {2015})}\BibitemShut {NoStop}%
\bibitem [{\citenamefont {Hassan}\ \emph {et~al.}(2017{\natexlab{b}})\citenamefont {Hassan}, \citenamefont {Galmiche}, \citenamefont {Harari}, \citenamefont {LiKamWa}, \citenamefont {Khajavikhan}, \citenamefont {Segev},\ and\ \citenamefont {Christodoulides}}]{hassan_chiral_2017}%
  \BibitemOpen
  \bibfield  {author} {\bibinfo {author} {\bibfnamefont {A.~U.}\ \bibnamefont {Hassan}}, \bibinfo {author} {\bibfnamefont {G.~L.}\ \bibnamefont {Galmiche}}, \bibinfo {author} {\bibfnamefont {G.}~\bibnamefont {Harari}}, \bibinfo {author} {\bibfnamefont {P.}~\bibnamefont {LiKamWa}}, \bibinfo {author} {\bibfnamefont {M.}~\bibnamefont {Khajavikhan}}, \bibinfo {author} {\bibfnamefont {M.}~\bibnamefont {Segev}}, \ and\ \bibinfo {author} {\bibfnamefont {D.~N.}\ \bibnamefont {Christodoulides}},\ }\bibfield  {title} {\enquote {\bibinfo {title} {Chiral state conversion without encircling an exceptional point},}\ }\href {\doibase 10.1103/PhysRevA.96.052129} {\bibfield  {journal} {\bibinfo  {journal} {Physical Review A}\ }\textbf {\bibinfo {volume} {96}},\ \bibinfo {pages} {052129} (\bibinfo {year} {2017}{\natexlab{b}})}\BibitemShut {NoStop}%
\bibitem [{\citenamefont {Nasari}\ \emph {et~al.}(2022)\citenamefont {Nasari}, \citenamefont {Lopez-Galmiche}, \citenamefont {Lopez-Aviles}, \citenamefont {Schumer}, \citenamefont {Hassan}, \citenamefont {Zhong}, \citenamefont {Rotter}, \citenamefont {LiKamWa}, \citenamefont {Christodoulides},\ and\ \citenamefont {Khajavikhan}}]{nasari_observation_2022}%
  \BibitemOpen
  \bibfield  {author} {\bibinfo {author} {\bibfnamefont {H.}~\bibnamefont {Nasari}}, \bibinfo {author} {\bibfnamefont {G.}~\bibnamefont {Lopez-Galmiche}}, \bibinfo {author} {\bibfnamefont {H.~E.}\ \bibnamefont {Lopez-Aviles}}, \bibinfo {author} {\bibfnamefont {A.}~\bibnamefont {Schumer}}, \bibinfo {author} {\bibfnamefont {A.~U.}\ \bibnamefont {Hassan}}, \bibinfo {author} {\bibfnamefont {Q.}~\bibnamefont {Zhong}}, \bibinfo {author} {\bibfnamefont {S.}~\bibnamefont {Rotter}}, \bibinfo {author} {\bibfnamefont {P.}~\bibnamefont {LiKamWa}}, \bibinfo {author} {\bibfnamefont {D.~N.}\ \bibnamefont {Christodoulides}}, \ and\ \bibinfo {author} {\bibfnamefont {M.}~\bibnamefont {Khajavikhan}},\ }\bibfield  {title} {{\selectlanguage {en}\enquote {\bibinfo {title} {Observation of chiral state transfer without encircling an exceptional point},}\ }}\href {\doibase 10.1038/s41586-022-04542-2} {\bibfield  {journal} {\bibinfo  {journal} {Nature}\ }\textbf {\bibinfo {volume} {605}},\ \bibinfo {pages} {256} (\bibinfo {year}
  {2022})}\BibitemShut {NoStop}%
\bibitem [{\citenamefont {Nye}(2023)}]{nye_universal_2023}%
  \BibitemOpen
  \bibfield  {author} {\bibinfo {author} {\bibfnamefont {N.~S.}\ \bibnamefont {Nye}},\ }\bibfield  {title} {\enquote {\bibinfo {title} {Universal state conversion in discrete and slowly varying non-{Hermitian} cyclic systems: {An} analytic proof and exactly solvable examples},}\ }\href {\doibase 10.1103/PhysRevResearch.5.033053} {\bibfield  {journal} {\bibinfo  {journal} {Physical Review Research}\ }\textbf {\bibinfo {volume} {5}} (\bibinfo {year} {2023}),\ 10.1103/PhysRevResearch.5.033053}\BibitemShut {NoStop}%
\bibitem [{\citenamefont {Nye}\ and\ \citenamefont {Kantartzis}(2024)}]{nye_adiabatic_2024}%
  \BibitemOpen
  \bibfield  {author} {\bibinfo {author} {\bibfnamefont {N.~S.}\ \bibnamefont {Nye}}\ and\ \bibinfo {author} {\bibfnamefont {N.~V.}\ \bibnamefont {Kantartzis}},\ }\bibfield  {title} {\enquote {\bibinfo {title} {Adiabatic state conversion for (a)cyclic non-{Hermitian} quantum {Hamiltonians} of generalized functional form},}\ }\href {\doibase 10.1063/5.0225403} {\bibfield  {journal} {\bibinfo  {journal} {APL Quantum}\ }\textbf {\bibinfo {volume} {1}},\ \bibinfo {pages} {046107} (\bibinfo {year} {2024})}\BibitemShut {NoStop}%
\bibitem [{\citenamefont {Kumar}\ \emph {et~al.}(2025)\citenamefont {Kumar}, \citenamefont {Gefen},\ and\ \citenamefont {Snizhko}}]{kumar_general_2025}%
  \BibitemOpen
  \bibfield  {author} {\bibinfo {author} {\bibfnamefont {P.}~\bibnamefont {Kumar}}, \bibinfo {author} {\bibfnamefont {Y.}~\bibnamefont {Gefen}}, \ and\ \bibinfo {author} {\bibfnamefont {K.}~\bibnamefont {Snizhko}},\ }\href {\doibase 10.48550/arXiv.2502.04214} {\enquote {\bibinfo {title} {General theory of slow non-{Hermitian} evolution},}\ } (\bibinfo {year} {2025}),\ \bibinfo {note} {arXiv:2502.04214 [quant-ph]}\BibitemShut {NoStop}%
\bibitem [{\citenamefont {Mertens}\ \emph {et~al.}(2021)\citenamefont {Mertens}, \citenamefont {Wesseling}, \citenamefont {Vercauteren}, \citenamefont {Corrales-Salazar},\ and\ \citenamefont {van Wezel}}]{mertens_inconsistency_2021}%
  \BibitemOpen
  \bibfield  {author} {\bibinfo {author} {\bibfnamefont {L.}~\bibnamefont {Mertens}}, \bibinfo {author} {\bibfnamefont {M.}~\bibnamefont {Wesseling}}, \bibinfo {author} {\bibfnamefont {N.}~\bibnamefont {Vercauteren}}, \bibinfo {author} {\bibfnamefont {A.}~\bibnamefont {Corrales-Salazar}}, \ and\ \bibinfo {author} {\bibfnamefont {J.}~\bibnamefont {van Wezel}},\ }\bibfield  {title} {\enquote {\bibinfo {title} {Inconsistency of linear dynamics and {Born}'s rule},}\ }\href {\doibase 10.1103/PhysRevA.104.052224} {\bibfield  {journal} {\bibinfo  {journal} {Physical Review A}\ }\textbf {\bibinfo {volume} {104}},\ \bibinfo {pages} {052224} (\bibinfo {year} {2021})}\BibitemShut {NoStop}%
\bibitem [{\citenamefont {Luitz}\ and\ \citenamefont {Piazza}(2019)}]{luitz_exceptional_2019}%
  \BibitemOpen
  \bibfield  {author} {\bibinfo {author} {\bibfnamefont {D.~J.}\ \bibnamefont {Luitz}}\ and\ \bibinfo {author} {\bibfnamefont {F.}~\bibnamefont {Piazza}},\ }\bibfield  {title} {{\selectlanguage {en}\enquote {\bibinfo {title} {Exceptional points and the topology of quantum many-body spectra},}\ }}\href {\doibase 10.1103/PhysRevResearch.1.033051} {\bibfield  {journal} {\bibinfo  {journal} {Physical Review Research}\ }\textbf {\bibinfo {volume} {1}},\ \bibinfo {pages} {033051} (\bibinfo {year} {2019})}\BibitemShut {NoStop}%
\bibitem [{\citenamefont {Žnidarič}(2022)}]{znidaric_solvable_2022}%
  \BibitemOpen
  \bibfield  {author} {\bibinfo {author} {\bibfnamefont {M.}~\bibnamefont {Žnidarič}},\ }\bibfield  {title} {{\selectlanguage {en}\enquote {\bibinfo {title} {Solvable non-{Hermitian} skin effect in many-body unitary dynamics},}\ }}\href {\doibase 10.1103/PhysRevResearch.4.033041} {\bibfield  {journal} {\bibinfo  {journal} {Physical Review Research}\ }\textbf {\bibinfo {volume} {4}},\ \bibinfo {pages} {033041} (\bibinfo {year} {2022})}\BibitemShut {NoStop}%
\bibitem [{\citenamefont {Martinez~Alvarez}\ \emph {et~al.}(2018)\citenamefont {Martinez~Alvarez}, \citenamefont {Barrios~Vargas},\ and\ \citenamefont {Foa~Torres}}]{martinez_alvarez_non-hermitian_2018}%
  \BibitemOpen
  \bibfield  {author} {\bibinfo {author} {\bibfnamefont {V.~M.}\ \bibnamefont {Martinez~Alvarez}}, \bibinfo {author} {\bibfnamefont {J.~E.}\ \bibnamefont {Barrios~Vargas}}, \ and\ \bibinfo {author} {\bibfnamefont {L.~E.~F.}\ \bibnamefont {Foa~Torres}},\ }\bibfield  {title} {\enquote {\bibinfo {title} {Non-{Hermitian} robust edge states in one dimension: {Anomalous} localization and eigenspace condensation at exceptional points},}\ }\href {\doibase 10.1103/PhysRevB.97.121401} {\bibfield  {journal} {\bibinfo  {journal} {Physical Review B}\ }\textbf {\bibinfo {volume} {97}},\ \bibinfo {pages} {121401} (\bibinfo {year} {2018})}\BibitemShut {NoStop}%
\bibitem [{\citenamefont {Yao}\ and\ \citenamefont {Wang}(2018)}]{yao_edge_2018}%
  \BibitemOpen
  \bibfield  {author} {\bibinfo {author} {\bibfnamefont {S.}~\bibnamefont {Yao}}\ and\ \bibinfo {author} {\bibfnamefont {Z.}~\bibnamefont {Wang}},\ }\bibfield  {title} {\enquote {\bibinfo {title} {Edge {States} and {Topological} {Invariants} of {Non}-{Hermitian} {Systems}},}\ }\href {\doibase 10.1103/PhysRevLett.121.086803} {\bibfield  {journal} {\bibinfo  {journal} {Physical Review Letters}\ }\textbf {\bibinfo {volume} {121}},\ \bibinfo {pages} {086803} (\bibinfo {year} {2018})}\BibitemShut {NoStop}%
\bibitem [{\citenamefont {Bohr}(2010)}]{bohr_atomic_2010}%
  \BibitemOpen
  \bibfield  {author} {\bibinfo {author} {\bibfnamefont {N.}~\bibnamefont {Bohr}},\ }\href {https://store.doverpublications.com/products/9780486479286} {{\selectlanguage {en}\emph {\bibinfo {title} {Atomic {Physics} and {Human} {Knowledge}}}}}\ (\bibinfo  {publisher} {Dover},\ \bibinfo {year} {2010})\BibitemShut {NoStop}%
\bibitem [{\citenamefont {Mermin}(2022)}]{mermin_there_2022}%
  \BibitemOpen
  \bibfield  {author} {\bibinfo {author} {\bibfnamefont {N.~D.}\ \bibnamefont {Mermin}},\ }\bibfield  {title} {\enquote {\bibinfo {title} {There is no quantum measurement problem},}\ }\href {\doibase 10.1063/PT.3.5027} {\bibfield  {journal} {\bibinfo  {journal} {Physics Today}\ }\textbf {\bibinfo {volume} {75}},\ \bibinfo {pages} {62} (\bibinfo {year} {2022})}\BibitemShut {NoStop}%
\bibitem [{\citenamefont {Schlosshauer}(2011)}]{schlosshauer_measurement_2011}%
  \BibitemOpen
  \bibfield  {author} {\bibinfo {author} {\bibfnamefont {M.}~\bibnamefont {Schlosshauer}},\ }\bibfield  {title} {{\selectlanguage {en}\enquote {\bibinfo {title} {The {Measurement} {Problem}},}\ }}in\ \href {\doibase 10.1007/978-3-642-20880-5_7} {{\selectlanguage {en}\emph {\bibinfo {booktitle} {Elegance and {Enigma}: {The} {Quantum} {Interviews}}}}},\ \bibinfo {editor} {edited by\ \bibinfo {editor} {\bibfnamefont {M.}~\bibnamefont {Schlosshauer}}}\ (\bibinfo  {publisher} {Springer},\ \bibinfo {address} {Berlin, Heidelberg},\ \bibinfo {year} {2011})\ pp.\ \bibinfo {pages} {141--160}\BibitemShut {NoStop}%
\end{thebibliography}

%
\end{document}